\documentclass[useAMS,usenatbib]{mn2e}
\usepackage{epsfig}
\usepackage{epstopdf}
\usepackage{graphicx}
\usepackage{amssymb}
\usepackage{amsmath}
\usepackage{graphicx}
\usepackage{subcaption}
\usepackage[utf8]{inputenc}
\usepackage[export]{adjustbox}
\usepackage{wrapfig}
\usepackage{textcomp}
\usepackage[usenames,dvipsnames]{color}
\usepackage{gensymb}
\usepackage{booktabs}
\usepackage{amsmath}
\usepackage{lscape}
\usepackage[english]{babel}
\usepackage{graphicx}
\usepackage{float}
\usepackage{fixltx2e}
\usepackage{soul}
\usepackage[normalem]{ulem}

\title[HOD of bright $z=6$ LBGs]{The environment and host haloes of the brightest $z\sim6$ Lyman-break galaxies}
\author[Peter Hatfield]{P. W. Hatfield$^{1}$\thanks{peter.hatfield@physics.ox.ac.uk}, R.A.A. Bowler$^{1}$, M.J. Jarvis$^{1, 2}$, C.L. Hale$^{1}$\\
$^{1}$Astrophysics, University of Oxford, Denys Wilkinson Building, Keble Road, Oxford, OX1 3RH, UK\\
$^{2}$Department of Physics, University of the Western Cape, Bellville 7535, South Africa\\
}

\begin{document}

\date{In original form 8th May 2015}

\pagerange{\pageref{firstpage}--\pageref{lastpage}} \pubyear{2015}

\maketitle

\label{firstpage}

\begin{abstract} By studying the large-scale structure of the bright high-redshift Lyman-break galaxy (LBG) population it is possible to gain an insight into the role of environment in galaxy formation physics in the early Universe. We measure the clustering of a sample of bright ($-22.7<M_{UV}<-21.125$) LBGs at $z\sim6$ and use a halo occupation distribution (HOD) model to measure their typical halo masses.  We find that the clustering amplitude and corresponding HOD fits suggests that these sources are highly biased ($b \sim 8$) objects in the densest regions of the high-redshift Universe. Coupled with the observed rapid evolution of the number density of these objects, our results suggest that the shape of high luminosity end of the luminosity function is related to feedback processes or dust obscuration in the early Universe - as opposed to a scenario where these sources are predominantly rare instances of the much more numerous $M_{UV} \sim -19$ population of galaxies caught in a particularly vigorous period of star formation. There is a slight tension between the number densities and clustering measurements, which we interpret this as a signal that a refinement of the model halo bias relation at high redshifts or the incorporation of quasi-linear effects may be needed for future attempts at modelling the clustering and number counts. Finally, the difference in number density between the fields (UltraVISTA has a surface density$\sim 1.8$ times greater than UDS) is shown to be consistent with the cosmic variance implied by the clustering measurements.

\end{abstract}

\begin{keywords}
galaxies: evolution -- galaxies: star-formation -- galaxies: high-redshift  -- techniques: photometric -- clustering -- LBGs
\end{keywords}

\section{Introduction}

\subsection{Lyman-break Galaxies}

The study of Lyman-break Galaxies (LBGs) is a long established probe of the high-redshift Universe (the first few billion years), with samples of many hundreds of star-forming galaxies now known to $z \sim 10$ (e.g. \citealp{Oesch2016,Bouwens2016,McLeod2016}). LBGs are particularly useful as it is possible to establish their photometric redshift to reasonable accuracy in a luminosity regime where spectroscopic confirmation is challenging (e.g. \citealp{Pentericci2014}). The neutral gas in the inter-galactic medium (IGM)  is essentially opaque to photons with wavelengths shorter than the `Lyman Break' (1216\AA , in the far ultraviolet). The source therefore appears faint bluewards of this wavelength, but retains its original luminosity redwards, creating a sharp drop in luminosity. When this spectrum is then redshifted, the location of the break provides a clear spectral feature with which to select galaxies to high-redshifts using broad-band filters.

The technique, originally developed in the early 1990's (\citealp{Guhathakurta1990,Steidel1992,Steidel1996}) in the context of $z \sim3$ galaxies, where the Lyman break is shifted into visible wavebands, first started providing large numbers of sources with the \textit{Hubble Space Telescope} (HST) in the late 1990's and 2000's (e.g. \citealp{Giavalisco2002,Bouwens2007,Dunlop2013}). More recently, the approach is being used to push scientific boundaries at $z\sim 6-9$ where the break is shifted into the near-infrared (see \citealp{Stark2016} for a review). Wide field surveys like the United Kingdom Infrared Telescope (UKIRT) Infrared Deep Sky Survey (UKIDSS, in particular the Ultra Deep Survey, UDS, \citealp{Hartley2013b}), and more recently public surveys on the Visible and Infrared Survey Telescope for Astronomy (VISTA) such as the UltraVISTA survey in the COSMOS field (\citealp{McCracken2012}) and the VISTA Deep Extragalactic Observations (VIDEO) survey (\citealp{Jarvis2013}) give access to the deep NIR images of the sky needed for detecting statistically significant samples of the brightest LBGs, which has lead to advances in the understanding of their star formation rates and number densities beyond the break in the luminosity function.

A key observable that can be calculated for LBG surveys (and galaxy surveys in general), is the luminosity or mass function, the comoving number density of galaxies as a function of absolute luminosity or stellar mass (see \citealp{Johnston2011} for a review). Measuring and understanding the evolution of luminosity functions with redshift allows us to trace the build-up and evolution of galaxies through cosmic time (\citealp{Madau2014}); is a key way to compare cosmological simulations of structure formation to observations (\citealp{Lacey2016,Clay2015}); and can be readily linked theoretically to the dark matter equivalent, the halo mass function (HMF; the comoving density of dark matter halos as a function of halo mass, see \citet{Murray2013} for a review of current constraints). Luminosity functions are typically observed to have the form of a Schechter function: $n(L)=\phi^{*}(L/L^{*})^{\alpha}\exp(-L/L^{*})$ (\citealp{Schechter1976}). In this parametrisation, $\alpha$ describes the power law behaviour of number density at the low-luminosity end, $L^{*}$ is the transition luminosity to the high luminosity exponential cutoff, and $\phi^{*}$ is a normalisation. The rest frame UV luminosity function for $z \sim 4-8$ has been determined by several studies (e.g. recent work by \citealp{McLure2013, Bouwens2015,Finkelstein2015,Bowler2015}) with broad agreement. The highest redshift constraints on the LBG luminosity function are currently at $z \sim 9-10$ e.g. \cite{Bouwens2015,Bouwens2016,McLeod2016}.

\subsection{Clustering}

Galaxies are formed and live in dark matter halos, and the environment of the host halo is believed to be of critical importance for the formation of the resident galaxies (\citealp{Cooray2002}). One way of obtaining information about the galaxy-halo connection is `abundance matching' - matching the galaxy comoving number density value to the halo mass that is predicted to have the same number density by theoretical considerations of the halo mass function/N-body simulations e.g. rarer galaxies are associated with more massive halos because such halos are rarer (\citealp{Vale2004}). Abundance matching however can only ever give an incomplete account of the connection due to three complications. Firstly, halos can host multiple galaxies, this can be partially mitigated through \textit{sub-}halo matching (\citealp{Moster2009}), but this assumes that the occupation statistics are the same for sub-halos and isolated halos. Secondly, scatter in the halo mass to galaxy mass/luminosity relation is not captured by abundance matching. Finally, variations in observational properties of a single population can bias results, in particular orientation or temporal effects e.g. if a given population has a different appearance 10 percent of the time, a straightforward abundance matching will erroneously place these sources with a different appearance in more massive halos as they are rarer, even though they are the same object as the underlying population. For this reason, other probes of the galaxy halo connection are needed. Information from lensing is very effective, either strong (e.g. \citealp{Paczynski1987,Soucail1988,Jullo2007}) or weak (galaxy-galaxy lensing, e.g. \citealp{Brainerd1995,Bartelmann1999,Coupon2015,Mandelbaum2013}), but requires the background sources to be at even higher redshift than the lenses, making it unfeasible for high ($z>2$) galaxies.

One viable and popular approach is to measure the clustering (2-point statistics) of the galaxies alongside the number counts (1-point statistics). This can then be linked to models/our theoretical understanding of structure formation to estimate the typical environment of the galaxies. One popular framework for modelling galaxy clustering is the `Halo Occupation Distribution' model (e.g. \citealp{Benson1999b,Ma2000,Seljak2000,Berlind2001,Zehavi2005}), which models the non-linear clustering of galaxies within individual halos, and the large scale clustering of the halos simultaneously, giving information about how many galaxies are in each halo as a function of halo mass. The HOD model has been applied extensively at $z=0$ (\citealp{Guo2015}) and $z=0.5-2$ (\citealp{McCracken2015,Coupon2015,Hatfield2016}) where large galaxy samples are available. In the more uncertain high-redshift regime, the HOD model has recently been applied to low-luminosity LBG galaxies at $z=4-7$ by \citet{Harikane2015} (where it has also been possible to compare to clustering predictions from hydrodynamical cosmological predictions e.g. the Bluetides simulation, see \citealp{Bhowmick2017}). It is crucial however to understand the relationship at the massive/most luminous end, as this is where AGN-driven feedback may have a role (\citealp{Croton2006,Bower2005,Silk2011}). There are preliminary hints that redshifts of $z=6-7$ may mark the onset of quenching (\citealp{Bowler2014,Bowler2015}), so this is a vital time period for galaxy evolution in the history of the Universe.

As well as being a crucial period for galaxy formation (see \citealp{Shapley2011} for a review), understanding large-scale structure/clustering at $z=5-8$ is also important for our understanding of \textit{reionization} (see \citealp{Zaroubi2013,Natarajan2014} for an observational and theoretical review respectively). The current best constraint on the average reionization redshift from the Planck mission  is $z=7.8-8.8$ (\citealp{PlanckCollaboration2016}), with many probes of the epoch (e.g. \citealp{Pentericci2014,Becker2015}) suggesting that some parts of the Universe could still be undergoing reionization by $z \sim 6$. \citet{McQuinn2007} suggest that differences in the clustering of LBGs and Lyman-$\alpha$ Emitting galaxies (LAEs) could give an insight into the possible `patchy' nature of reionization.  The Lyman-alpha line is suppressed if the source is in a largely neutral region which biases the observations of LAEs towards large ionized H{\sc ii} regions. The result is a larger `observed' clustering for LAEs than the `intrinsic' clustering of the underlying objects - effectively neutral regions obstruct the line of sight in a way that enhances the clustering of LAEs. LBGs however do not receive such an effect on their clustering, so a boost in the clustering of LAEs relative to LBGs (properly controlling for other variables) could be indicative of reionization. Such an effect is yet to be conclusively measured, e.g. \citet{Ouchi2010} find little evidence at $z=6.6$ with 207 LAEs observed with the Subaru telescope, but this approach and others like it are likely to give improved constraints to the nature of the epoch of reionization over the coming years.

\subsection{Objectives of this work}

LBG studies can be informally divided into analyses of `faint' galaxies (in extremely deep, but narrow surveys), and `bright' galaxies (in slightly less deep, but extremely wide surveys). \citet{Harikane2015} provide an analysis of the clustering of relatively faint LBGs found within HST deep surveys at $z = 4-7$. In this study we seek to extend these measurements to brighter luminosities by utilising wider area surveys. To do this, we measure and model the clustering of the \citet{Bowler2015} high luminosity $z \sim 6$ sample, which covers the degree-scale UDS and UltraVISTA fields. A clustering analysis of a subset of the UDS sample has been performed in \citet{McLure2009}, who modelled the correlation function with a single power law, concluding the sources are in dark matter haloes of masses $10^{11.5-12} M_{\odot}$. In this study we perform a similar analysis, but extend to a full HOD model, including an additional field and using deeper data available. Using this enlarged sample, we are then able to discuss what our results mean for feedback processes, models of structure formation, and cosmic variance at high redshift. While samples of bright galaxies do exist at $z>6.5$ (\citealp{Bowler2015}), they are too small to provide constraints on the clustering, and hence we limit our analysis to $z\sim6$.

The structure of this paper is as follows. In Section 2 we describe the sample of LBGs used in this study. In Section 3 we discuss how we measured the correlation function in the sample and constructed our halo occupation distribution models and fitting process. The results are presented in Section 4. In Section 5 we discuss our results, linking them to the literature, and interpreting the cosmic variance between the fields in light of our measurements. Magnitudes are given in the AB system (\citealp{Oke1983}) and all calculations are in the concordance cosmology $\sigma_{8}=0.8$, $\Omega_{\Lambda}=0.7$, $\Omega_{m}=0.3$ and $H_{0}=70 \text{ km} \text{ s}^{-1} \text{Mpc}^{-1}$ unless otherwise stated.

\section{Data and Sample Selection} \label{sec:Data}

In this study we use the high luminosity Lyman break galaxy sample of \citet{Bowler2015}. Deep optical and infrared data (spanning wavelengths of $0.3-2.5 \mu$m) across two main fields (see Fig. \ref{fig:field_geometry}) was used to select the sample; we summarise the observations and selection criteria below, but see \citet{Bowler2015} for a more in depth description.

\subsection{UltraVISTA/COSMOS} \label{sec:UltraVISTA}

UltraVISTA (\citealp{McCracken2012,Laigle2016}) is the deepest of the 6 public surveys on the VISTA telescope, providing $YJHK_{s}$ near infra-red data covering the Cosmic Evolution Survey (COSMOS) field (\citealp{Scoville2007}). The `paw-print' focal plane of VISTA and the survey observing strategy give a continuous `deep' field, with discontinuous `ultra-deep' stripes across it that receive more observing time. \citet{Bowler2015} also used $u^{*}$, $g$, $r$ and $i$ optical data from the T0007 release of CFHTLS in the D2 field, as well as Subaru/SuprimeCam $z'$-band imaging. The maximal area of overlap of these datasets is in the one square degree of CFHTLS, of which 0.62 deg$^2$ has ultra-deep UltraVISTA data, and 0.38 deg$^2$ shallower UltraVISTA. The majority of the sample is in the ultra-deep field and hence for our purposes here we only use the ultra-deep 0.62 deg$^2$ (see Fig. \ref{fig:field_geometry}).

\subsection{UDS} \label{sec:UDS}

\begin{figure*}
\includegraphics[scale=0.3]{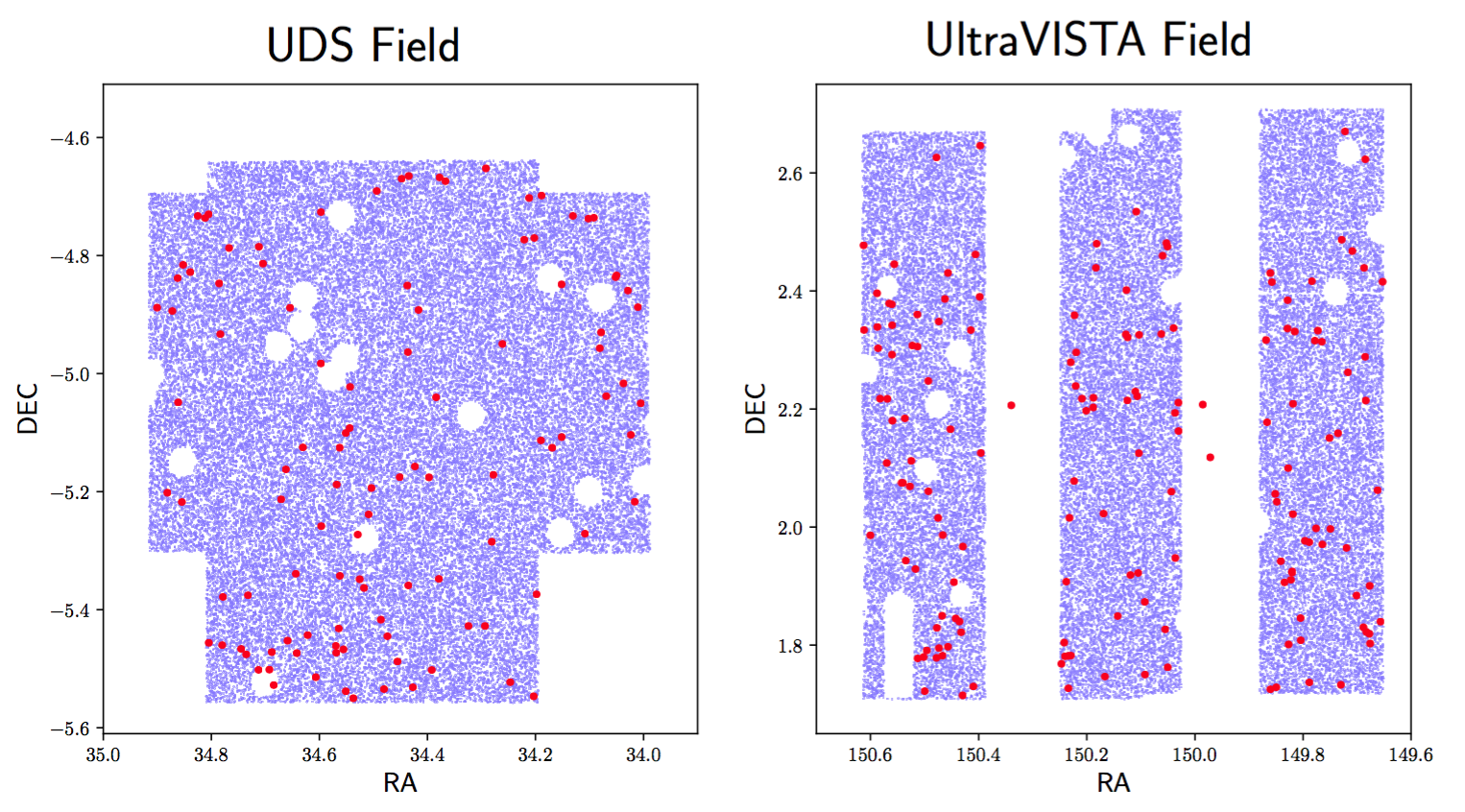} 
\caption{The geometry of the UDS and UltraVISTA fields. The red points are the galaxy locations from \citet{Bowler2015}. The blue points are the random points chosen to cover the fields used for the construction of $RR$ for the calculation of the correlation function. The three galaxies in the UltraVISTA field that are not surrounded by blue points are the $z=6$ sources detected in the deep (as opposed to ultra-deep) part of the UltraVISTA field, that we do not include in this study. The overall shape of the fields is predominantly determined by the part of the sky that the multiple different surveys overlap in. The small scale gaps and holes are foreground stars and detector artefacts etc. See figures 1 and 2 of \citet{Bowler2014} to see how the irregular footprints arise from the intersection of the sky patches covered by different surveys.}
\label{fig:field_geometry}
\end{figure*}

For the UKIDSS UDS field, \citet{Bowler2015} used $B$, $V$, $R$, $i$ and $z'$ data from the Subaru XMM-Newton Deep Survey (SXDS, \citealp{Furusawa2008}), and $J$, $H$ and $K$ band data from the DR10 of the UKIDSS UDS (\citealp{Lawrence2006}). Again separate $z'$-band data from Subaru/SuprimeCam was obtained, and in addition, $Y$ band data from the VIDEO survey (\citealp{Jarvis2013}) was also used. The total overlapping area is 0.74 deg$^2$ (see Fig. \ref{fig:field_geometry}).

\subsection{Candidate Selection} \label{sec:selection}

Again, \cite{Bowler2015} describes the full sample selection, but we summarise the process here. Sources were detected with {\sc SExtractor} (\citealp{Bertin1996}), and photometric redshifts were determined with {\sc LePhare} (\citealp{Arnouts1999,Ilbert2006}). Contaminant populations (low redshift interlopers and brown dwarfs) were removed in the SED fitting process. This leaves 156 and 107 $5.5<z <6.6$ galaxies in the UltraVISTA and UDS fields respectively. The UltraVISTA field was found to have a higher surface density than the UDS field (by a factor of $\sim 1.8$); potential causes for this, including lensing and cosmic variance are discussed in section 7 of \citet{Bowler2015}.

This process gives in total 263 LBGs in the range $-22.7< M_{UV} < -20.5$ with $5.5<z<6.5$ over 1.35 deg$^2$. We take our final sample as all 161 sources with $M_{UV} <-21.125$, as the sample completeness drops off rapidly faintwards of this value, as discussed in \cite{Bowler2015}, see their figure 6, but is fairly constant with magnitude brightwards of this value.

\section{Correlation Functions and HOD Modelling} \label{sec:corr_functions}

There is a large selection of statistical measurements that can be used to characterise the clustering of extragalactic sources and large-scale structure, including nearest neighbour \citep{Bahcall1983}, genus \citep{Gott2008a}, power spectrum \citep{Tegmark2003} and counts in cells \citep{White1979}. In this study we measure and model the two-point correlation function, the excess probability of how much more likely two galaxies are to be at a given separation than a random uniform distribution (this statistic can be linked to other measurements e.g. counts in cells statistics are `averaged' correlation functions, and the power spectrum is the Fourier transform of the correlation function). 

The underlying meaningful physical relation is the full three dimensional spatial correlation function; however we only have the observables of angular separations and relatively coarse redshift information. Limber Inversion (\citealt{Limber:1954zz}) provides a way to connect the two - we calculate a model spatial correlation function, and project to an angular correlation function.

\subsection{The Angular Correlation Function}\label{sec:ACF_def}

The angular two-point correlation function $\omega(\theta)$ is defined by:

\begin{equation}
dP=\sigma(1+\omega(\theta))d\Omega ,
\end{equation}

where $dP$ is the probability of finding two galaxies at an angular separation $\theta$, $\sigma$ is the surface number density of galaxies, $d\Omega$ is the solid angle. This is most commonly estimated by calculating $DD(\theta)$, the normalised number of galaxies at a given separation in the real data, and $RR(\theta)$, the corresponding figure for a synthetic catalogue of random galaxies identical to the data catalogue in every way (i.e. occupying the same field) except position. We use the \citet{Landy1993} estimator:

\begin{equation}
\omega(\theta)=\frac{DD-2DR+RR}{RR} ,
\end{equation}

which also uses $DR(\theta)$,  data to random pairs, as it has a lower variance (as an estimator) and takes better account of edge effects.

Uncertainty on measurements of the correlation function can be broken down into i) `Poisson-like' uncertainty, which depends on the number density of the population tracing the underlying density field, and ii) `cosmic variance' uncertainty, which depends on how representative the volume under consideration is of structures on the scale of interest (see also section \ref{sec:cosmic_variance}). Estimates of uncertainty can be either `internal' (based solely on the data itself), `external' (based on comparison with mock catalogues from simulations or similar) or analytic (based on analytic equations from the the theory of clustering statistics, typically using 3rd or 4th order statistics). In this work uncertainties are calculated with the internal `bootstrap-object resampling' method, which samples the individual galaxies with replacement from the dataset, from which we recalculate the correlation function (see \citealp{Ling1986}). Repetition of this process produces multiple `realisations' of the correlation function, from which the covariance matrix of the $\omega(\theta)$ values can be estimated. In terms of `Poisson-like' uncertainty, it is possible to calculate the error bars from Poisson uncertainty on the $DD$ values, but \citet{Cress1996} and \citet{Lindsay2014} found errors calculated in this manner were a factor of 1.5 to 2 smaller than those estimated with bootstrap-object resampling. It is particularly important to account for covariance between adjacent angular space bins in the small-number counts regime here as each galaxy will contribute to multiple bins. In this paper we use 100 bootstrap resamplings to estimate the uncertainty at the 16th and 84th percentiles.

The other important internal approaches to estimating the uncertainty on measurements of the correlation function involve dividing the field into $N$ sub-regions. `Bootstrap-volume' resampling (e.g. \citealp{Maddox1990}) then samples $M$ of these sub-regions (with replacement) repeatedly, calculating the correlation function each time, similarly to bootstrap-object resampling. The `jackknife' method (\citealp{Shao1986}) measures how the estimate of the correlation function varies when individual sub-regions are removed. Bootstrap-object estimates are known to under-estimate the true cosmic variance uncertainty e.g. \citet{Fisher1994}. \citet{Mo1992} found that analytic and external mock estimates of uncertainty agreed well on all scales, but that bootstrap-object estimates only agreed with analytic estimates on small scales, underestimating the true uncertainty by $\sim35$ percent on scales comparable to the size of the field. Volume resampling methods conversely can slightly overestimate the true uncertainty; \citet{Norberg2008} discuss the advantages and disadvantages of bootstrap-volume and jackknife, finding that both overestimate the size of error bars by about $\sim$40 percent on scales relevant to this work when compared to external estimates based on mocks (which are beyond the scope of this work, but are typically considered to be closer to the `true' uncertainty). In terms of sub-dividing a survey into smaller volumes, \citet{Cabre2007} note that irregular shapes can jeopardise the estimation of correlation function uncertainty estimates for internal methods, and any subdivision of our survey geometry would certainly have very irregular shapes. In summary, bootstrap-object uncertainty estimates capture `Poisson-like' uncertainty well, but imperfectly captures `cosmic variance' uncertainty compared to bootstrap-volume methods. However given the difficulties of doing volume resampling with an irregular field geometry, and, perhaps most importantly, that our correlation function is calculated treating two separate highly independent fields as one, mitigating `cosmic variance' uncertainty, we conclude that bootstrap-object uncertainty estimates could slightly underestimate the error bars on our measurements of the correlation function on large scales, but are likely sufficiently accurate for our purposes.

For the construction of our random catalogue we created a mask over the fields to exclude image artefacts and foreground stars. Five galaxies in the UDS field were found to be within the masked area. Although the mask may be a little conservative, it is likely that our measurements of clustering in the vicinity of these sources will be heavily biased by the presence of the artefact being masked, so we do not use these five galaxies when calculating the correlation function (although it makes very little difference to our analysis). The fact that the survey area has a finite area gives a negative offset to the true correlation function, usually known as the integral constraint. As per \citet{Beutler2011} and \citet{Hatfield2016}, we calculate the integral constraint using the numerical approximation of \citet{Roche1999} and treat it as part of the model when fitting parameters. In this paper we calculate the correlation function with both the binning method ($DD$ and $RR$ are how many galaxy pair separations in each angular scale bin) and the continuous estimation/kernel smoothing method described in \citet{Hatfield2016}.

\subsection{Halo Occupation Distribution modelling}\label{sec:HOD_def}

Halo Occupation Modelling is an increasingly popular way of modelling galaxy clustering measurements. We do not describe the full details of the scheme here, see \citet{Coupon2012} and \citet{McCracken2015} for a more complete breakdown. A given set of galaxy occupation statistics is given, usually parametrised by 3-5 numbers, e.g. the number of galaxies in a halo as a function of halo mass. The model correlation function is broken down to a `1-halo' term, describing the small-scale clustering of galaxies within an individual halo, and a `2-halo' term, describing the clustering of the halos themselves. The `1-halo' term is constructed by convolving the profile of galaxies within a halo with itself, weighting by the number of galaxies in the halo, and then integrating over all halo masses. The profile is usually taken to be one galaxy at the centre of the halo (the `central') and all other galaxies tracing a Navarro-Frank-White (NFW; \citealp{Navarro1996}) profile. The 2-halo term is constructed by scaling the dark matter linear correlation function by the weighted-average halo bias of the host halos.

The most general HOD parametrisation commonly used is that of \citet{Zheng2005}, that gives the total number of galaxies in a halo as:

\begin{equation}  \label{eq:params_tot}
\langle N_{tot}(M_{h}) \rangle=\langle N_{cen}(M_{h}) \rangle+\langle N_{sat}(M_{h}) \rangle ,
\end{equation}

the total number of central galaxies as:

\begin{equation}  \label{eq:params_cen}
\langle N_{cen}(M_{h}) \rangle=\frac{1}{2} \left(1+ \mathrm{erf} \left( \frac{\log_{10}M_{h}-\log_{10}M_{min}}{\sigma_{\log_{10} M}} \right)  \right) ,
\end{equation}

and the total number of satellites as:

\begin{equation}  \label{eq:params_sat}
\langle N_{sat}(M_{h}) \rangle=\langle N_{cen}(M_{h}) \rangle \left( \frac{M_{h}-M_{0}}{M_{1}} \right)^{\alpha} .
\end{equation}

This model has five parameters; $M_{min}$ describes the minimum halo mass required to host a central galaxy, $\sigma_{\log M}$ describes how sharp this step jump is (equivalently to the central to halo mass scatter), $M_{0}$ is a halo mass below which no satellites are found, and $M_{1}$ is the scale mass for accumulating satellites ($M_{0}$ is typically a lot smaller than $M_{1}$, so $M_{1}$ is commonly said to be the halo mass at which the first satellite is accreted, although analytically they are very slightly different - this is the difference between $M_{1}$ and $M'_{1}$ used by some authors). The power law index $\alpha$ describes how the number of satellites grows with halo mass. Although we have the largest sample of bright LBGs at these redshifts, this is still only a comparatively small sample for HOD modelling. Thus, in order to reduce the number of parameters in the model (six once duty cycle is included, see Section \ref{sec:DC}), we fix some as functions of others.

As per \citet{Harikane2015} we fix $\sigma_{\log M}=0.2$. The assumptions that go into this choice however are based on results at much lower redshifts (\citealp{Kravtsov2004,Zheng2005,Conroy2006}) which do not necessarily hold at these early times, when the luminosity-halo scatter is fairly unconstrained. Indeed \citet{Hatfield2016} found a scatter of $\sim0.6$ consistent with the data at $z\sim1$. However fortunately for our purposes (unfortunately from the perspective of using clustering to infer the scatter) the 2-point statistics have very little dependence on the scatter. Hence our conclusions do not alter dramatically with choice of $\sigma_{\log M}$, and so we fix it as the same as the \citet{Harikane2015} value for ease of comparison.

We additionally fix $\alpha=1$; this is both the fiducial value (it is logical to expect that once in the most massive halo regime that the number of satellites scales linearly with the halo mass, as the bulk of the halo mass will have been accreted), as well as the result found by most measurements at moderate ($z<2$) redshift (e.g. \citealp{Hatfield2016,Coupon2012}).

We investigate the consequences of allowing various parameters to be fixed or free in the fitting process. As per equations 54 and 55 in \citet{Harikane2015}, if not free, $M_1$ and $M_0$ are fixed as functions of $M_{min}$ following the $z=0-5$ results of \citet{Conroy2006}:

\begin{equation}  \label{eq:M1}
\log{{(M_{1}/M_{\sun})}}=1.18 \log{(M_{min}/M_{\sun})}-1.28 ,
\end{equation}

\begin{equation}  \label{eq:chi}
\log{(M_{0}/M_{\sun})}=0.76 \log{(M_{1}/M_{\sun})} +2.3 .
\end{equation}

In this work we use the halo mass function of \cite{Behroozi2012}, and the halo bias of \citet{Tinker2010}.

\subsection{Duty Cycle} \label{sec:DC}

The role of a duty cycle (DC) is the main difference to be incorporated when modelling LBG galaxies at high redshift compared with studies in the local Universe. Clustering analyses of LBGs typically find that there is a mismatch between the measured number density, and the number density implied by the clustering (\citealp{Ouchi2010}). This is in agreement with current understanding of galaxies at these redshifts that suggest that star formation may be highly episodic e.g. \cite{Stark2009}. Typically the occupation statistics model implied by fitting \textit{only} to the clustering will suggest a larger comoving number density than is observed in the luminosity function. This discrepancy is typically explained by invoking a \textit{duty cycle}, that the observed LBGs have luminosities that vary dramatically in time, and are being observed only when in a bright phase. This illustrates the importance of understanding clustering alongside the number counts. With a duty cycle of 10 percent (i.e. it is only in its bright phase 10 percent of the time), the underlying galaxy appears 10 times rarer than it actually is. A straight abundance matching in this scenario would then mistakenly put them in rarer, and thus more massive, halos.

Without incorporating the duty cycle, the implied comoving number density is the mean number of galaxies in a halo, times the halo mass function, integrated over all halo masses. This number is then multiplied by the duty cycle to give the model comoving density:

\begin{equation}  \label{eq:n_gal}
n_{gal}=\textrm{DC} \times \int^{\infty}_{0} \textrm{HMF}(M_{h}) \times \langle N_{tot}(M_{h}) \rangle \mathrm{d}M_{h} , 
\end{equation}

where HMF is the halo mass function and DC is the duty cycle.

\subsection{MCMC Fitting} \label{sec:MCMC_description}

To compare with observations, we use the {\sc Halomod} \footnote{https://github.com/steven-murray/halomod} code (Murray, Power, Robotham, in prep.) to calculate the spatial correlation function. We then project this to an angular correlation function (as per \citealt{Limber:1954zz}, using a redshift distribution derived by smoothing the point estimates of the LBG redshifts), and subtract off the numerical approximation of the integral constraint to get our final model correlation function.

We use {\sc Emcee}  \footnote{http://dan.iel.fm/emcee/current/} (\citealp{Foreman-Mackey2012})  to provide a Markov Chain Monte Carlo sampling of the parameter space to fit our correlation function. We use a likelihood of:

\begin{equation}  \label{eq:chi}
\chi^{2}= \frac{[\log n_{\mathrm{\mathrm{gal}}}^{\mathrm{obs}}-\log n_{\mathrm{\mathrm{gal}}}^{\mathrm{model}}]^{2}}{\sigma_{\log n}^{2}}
\end{equation}
\begin{equation}  \label{eq:chi}
+  \sum\limits_{i,j} [\omega^{\mathrm{obs}}(\theta_{i})-\omega^{\mathrm{model}}(\theta_{i})] [C^{-1}_{i,j}][\omega^{\mathrm{obs}}(\theta_{j})-\omega^{\mathrm{model}}(\theta_{j})] , \nonumber
\end{equation}

where $n_{\mathrm{\mathrm{gal}}}^{\mathrm{obs}}$ is the observed galaxy number density, $n_{\mathrm{\mathrm{gal}}}^{\mathrm{model}}$ is the model galaxy number density, $\sigma_{\log n}$ is the error on the log of the  number density including both Poisson noise and cosmic variance (calculated as per \citealt{Trenti2008}), $\theta_{i}$ are the angular scales we fit over, $\omega^{\mathrm{obs}}$ is the observed angular correlation function, $\omega^{\mathrm{model}}$ is the angular correlation function of a given model, and $C_{i,j}$ is the covariance matrix of the measurements of the correlation function from the bootstrapping.

In addition to the uncertainty on the number counts and the covariance of the clustering measurements, there is in general \textit{``cosmic covariance''} between observed number counts and measured clustering amplitude. For an over-dense region of the Universe, typically the number counts and measured clustering amplitude will both be higher than the global value, and vice versa for an under-dense region (e.g. see \citealp{Lacasa2016}). This covariance would typically lead to an underestimate of the true uncertainties on our inferred HOD parameters. However, the fact that our main analysis is performed jointly over two highly independent fields (one of which is known to be over-dense and the other under-dense) mitigates the impact of this effect; we conclude not accounting for this covariance is unlikely to affect our key results.

When the parameters are free, we use a uniform prior over $10<\log_{10}{(M_{\mathrm{min}}/M_{\odot})}<13$, $\log_{10}{(M_{\mathrm{min}}/M_{\odot})}<\log_{10}{(M_{\mathrm{1}}/M_{\odot})}<14$ (uniform in log space) and $0<\textrm{DC}<1$. We used 20 walkers with 1000 steps, which have starting positions drawn uniformly from the prior. When we fit the bias $b$ directly (Model E in section \ref{sec:full_sample_modelling}) we use a uniform prior over $1<b<30$.

We use 500,000 random data points in this study. As per \citet{Hatfield2016} and \citet{Hatfield2016a}, we use 100 bootstrap-object resamplings to estimate the uncertainty at the 16th and 84th percentiles of the resampling (corresponding to plus/minus one standard deviation). For $n_{\mathrm{\mathrm{gal}}}^{\mathrm{obs}}$, we use the value obtained when integrating the luminosity function brightwards to infinity from $M_{UV}=-21.125$ (as opposed to the number obtained by dividing the number of sources by the volume probed) as the luminosity function already has incompleteness factored in. This equates to $n_{\mathrm{\mathrm{gal}}}^{\mathrm{obs}}=4.1 \times 10^{-5}$Mpc$^{-3}$, for the fields combined. This is from the best-fitting double power law model in \cite{Bowler2015}. Using the best-fitting Schecter function gives the marginally lower value of $n_{\mathrm{\mathrm{gal}}}^{\mathrm{obs}}=3.8 \times 10^{-5}$Mpc$^{-3}$. Changing from one value to the other does not impact our conclusions. The main sources of incompleteness are blending with foreground sources, and misclassification of true $z \sim 6$ LBGs as dwarf stars or lower redshift contaminants, see \cite{Bowler2015}.

\section{Results} \label{sec:RESULTS}

\subsection{Clustering Measurements} \label{sec:full_sample_measurements}

\begin{figure*} 
\includegraphics[scale=0.8]{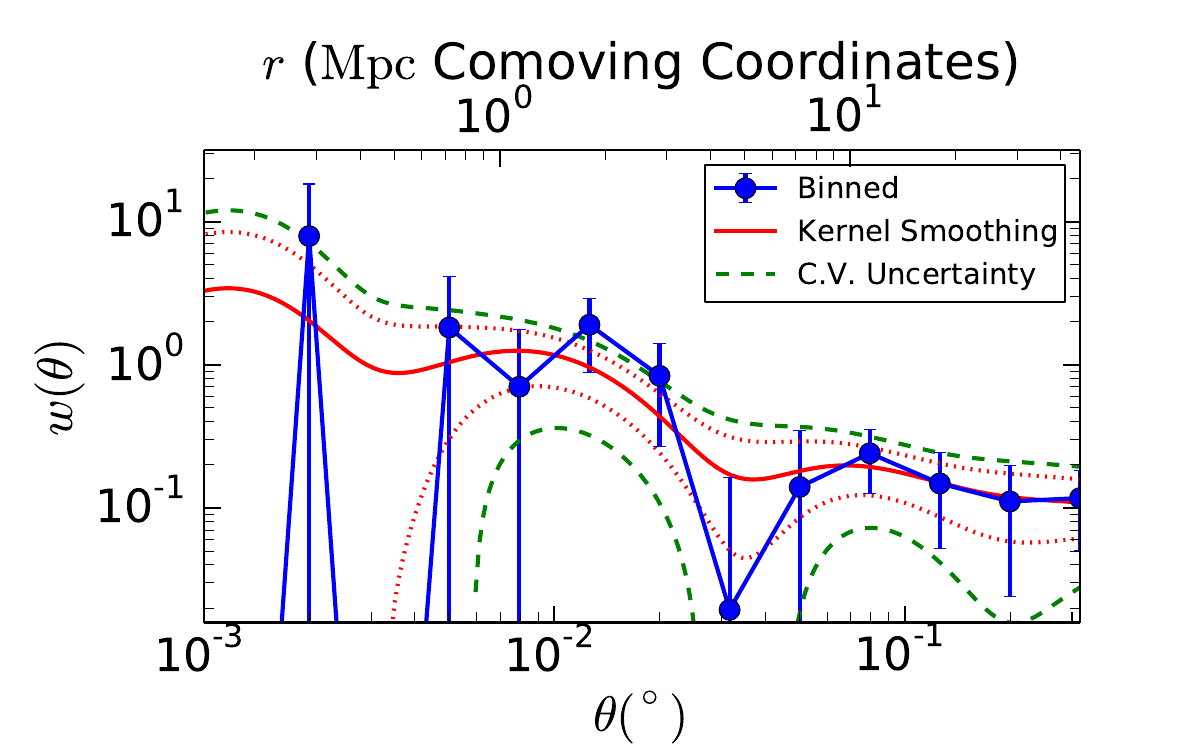}
\caption{The angular correlation function for our sample of bright ($M_{UV}<-21.125$) $z\sim6$ LBGs from Bowler et al., (2015). The figure shows the correlation function estimated both with a binning method (blue points), and a kernel smoothing method (red curve, with dotted lines showing the error on the measurements). The error bars on the measurements from both the binning method and the kernel smoothing method are calculated from the 16th and 84th percentiles of the set of bootstrapped measurements. The dashed green curves show the `cosmic variance corrected' uncertainty on the kernel smoothed clustering measurements, used and discussed in section \ref{sec:clustering_cosmic_variance}. }
\label{fig:full_acf}
\end{figure*}

Fig. \ref{fig:full_acf} shows the angular correlation function of the full sample over the range $10^{-3 }<\theta / \textrm{deg}<10^{-0.5} $, estimated with both the binning approach (where galaxy pair separations are counted in discrete angular ranges) and the kernel smoothing method (where the distribution of galaxy pair separations is smoothed to produce a continuous estimation of the correlation function in the angular range under consideration). They (as expected) agree well, and produce the familiar approximate power law $\sim \theta^{-0.8}$, although the kernel smoothing method is able to cope better with bins that contain a small number of pairs.  For the rest of our analysis, we take the value of the smoothed correlation function, at the ten angular scales calculated for the bins, as our final measurements \footnote{With the continuous estimation of the correlation function, one can in principle extract an estimate of the correlation function at an arbitrary number of angular scales in the range probed. However this gives dramatically diminishing returns as adjacent measurements would be increasingly covariant e.g. one could take the estimate of the correlation function at 1000 points in the angular range for which we estimate the correlation function, but adjacent points would be almost perfectly correlated and no extra information would be gained.}. Fig. \ref{fig:full_acf} also shows the estimate of the cosmic variance on the clustering measurements used in section \ref{sec:clustering_cosmic_variance}.

In general, measurements of clustering at different scales will be covariant as individual galaxies contribute multiple times to $DD$, usually at different scales. Furthermore, extra care with covariances is needed when using the kernel method, as a given galaxy pair contributes at a range of scales (this can be mitigated by picking measurements larger than the smoothing scale, but is important to keep track of here as we are in a low-data regime). Therefore we also construct the covariance matrix from the bootstrapped samples for our measurements, in order to account for these covariances in the fitting process. We show the correlation matrix (the covariance matrix with each value normalised by the standard deviation of each measurement) in Fig. \ref{fig:covariance}. Not taking covariances into account would be the equivalent of ignoring the off-diagonal values, which are non-negligible, particular at the large and very small scales.

\begin{figure} \label{fig:covariance}
\includegraphics[scale=0.5]{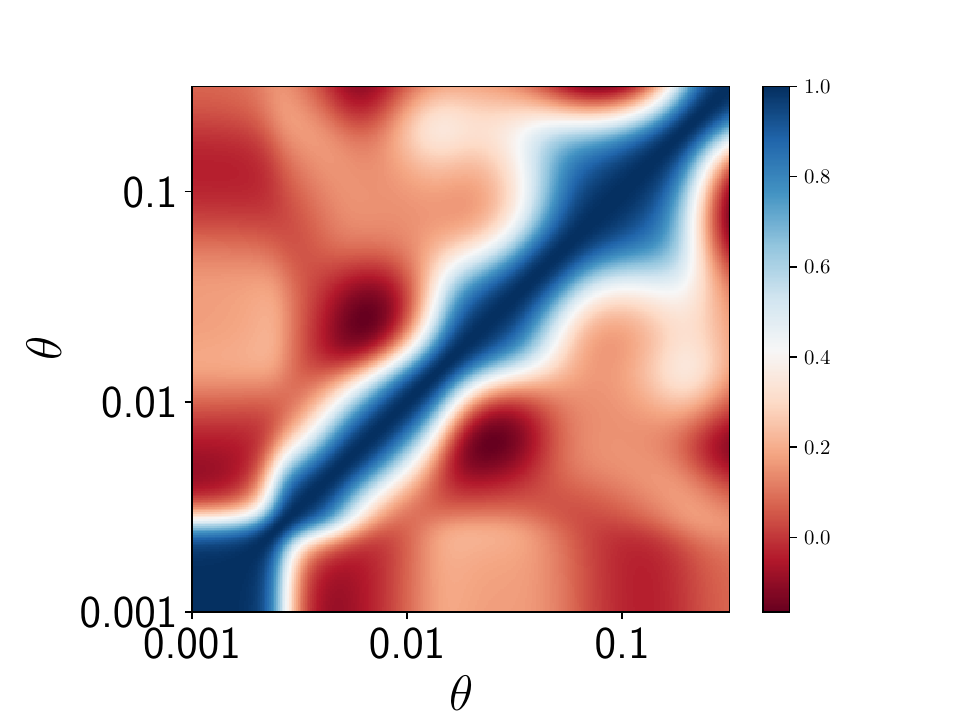}
\caption{The correlation coefficients (covariance normalised) of our measurements. Blue values are positive correlations, red values are negative values. The blue diagonal corresponds to the standard deviation measurements (a random variable is always perfectly correlated with itself). }
\label{fig:covariance}
\end{figure}

\begin{figure} 
\includegraphics[scale=0.45]{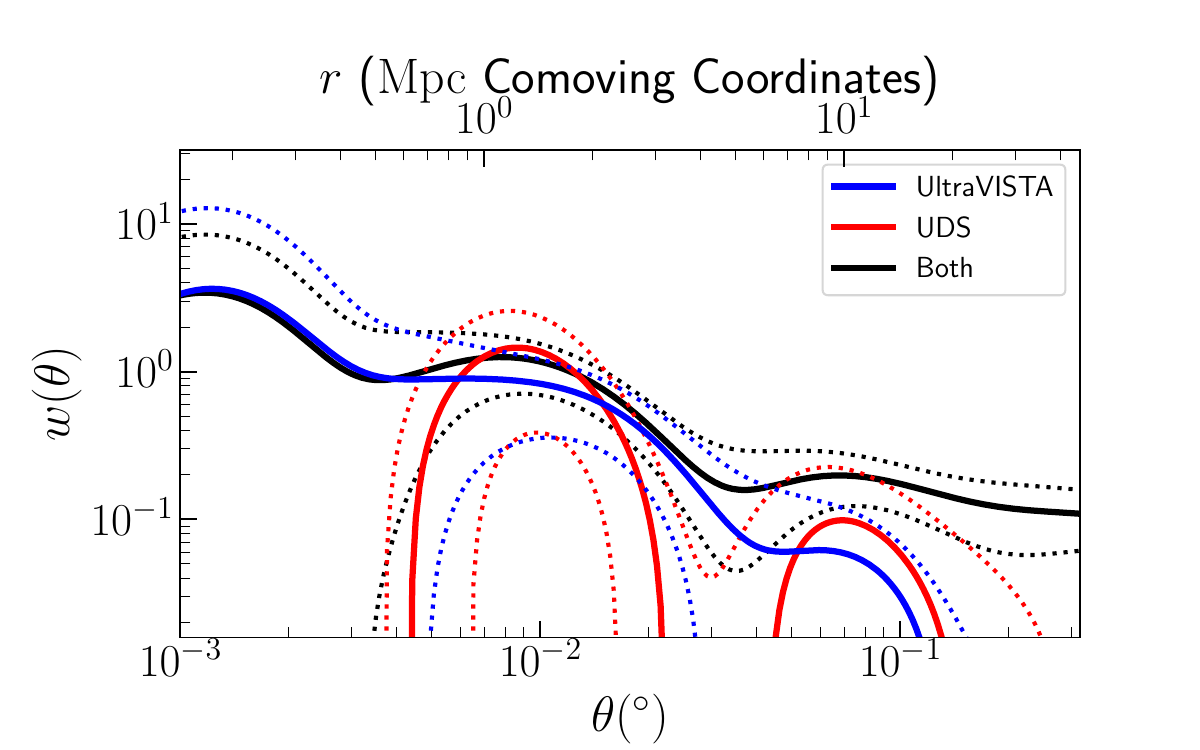}
\caption{The angular correlation function of our LBG sample, calculated for the UltraVISTA field (blue), the UDS field (red), and the fields collectively (black).}
\label{fig:full_acf_both}
\end{figure}

We also calculate the correlation function separately for the UDS and UltraVISTA fields individually, shown in Fig. \ref{fig:full_acf_both}. The single field correlation functions were calculated based solely on each individual field - random data points and $RR$ are re-calculated for each sub-field, $DD$ and $DR$ normalisation is based just on the galaxy number counts for the separate fields etc. Small number statistics make the correlation function for each field extremely noisy, particularly for the UDS field where we only have 64 LBGs with $M_{UV}<-21.125$ in the sample. The error bars are smaller for the measured correlation function when both fields are used, than for the individual fields, as expected. It is clear that all small scale pairs are from the UltraVISTA field - the UDS correlation function has no power on scales smaller than $\sim 10^{-2.5}$ deg\footnote{This illustrates a weakness of the random catalogue approach to estimating the correlation function - when zero pairs are observed, this approach outputs $\omega=-1$ for every bootstrapped catalogue (as zero pairs can not be bootstrapped into a non-zero number of pairs), as opposed to more realistically giving no constraints.}. On scales of $\sim 10^{-2}-10^{-1}$ deg the amplitudes of the UDS and UltraVISTA correlation functions are roughly consistent with each other, and with the correlation function calculated for the fields together\footnote{The UDS field correlation function has a `dip' relative to the UltraVISTA and `Both' correlation functions at about $0.02-0.04 ^{\circ}$. We are not aware of any systematic (instrumental or in observing strategy) that inhibits finding galaxy pairs at this angular separation, so believe this is likely a $\sim$$1.5$-$\sigma$ chance fluctuation, as opposed to a significant discrepancy.}. Finally on scales $>10^{-1}$ deg, the correlation function is moderately larger for the fields considered together than for the fields considered individually. This is because with just one field, essentially there is only access to $\sim$one large mode of fluctuation, so variance in the large-scale modes can not be observed. However with more than one field these large-scale variations can now be measured. Mathematically this behaviour can be seen as a result of the use of the \citet{Landy1993} estimator - $DR$ is moderately large on the large scales ($\gtrsim 0.1$ deg) for each individual field, but is smaller when the fields are considered collectively, as more of the randomly placed `dummy' galaxy points are in the UDS field but more of the real galaxy points points are in UltraVISTA. The estimate of the correlation function on larger scales when using the earlier \citet{Peebles1974} estimator $\omega=\frac{DD}{RR}-1$ is much lower and essentially is the weighted average of the measurements from each individual field (suggesting that the value of $DR$ is critical). Equivalently, the integral constraint (which starts to become significant on scales comparable to the maximum length scale accessed by a survey) is smaller when the fields are considered together, as there are large numbers of dummy-dummy pairs with one member of the pair in one field, and one in the other field, which contribute to the normalising denominator, but not the numerator, of the expression for the integral constraint in \citet{Roche1999} that we use in this work.

\subsection{Modelling Results} \label{sec:full_sample_modelling}

We carry out several MCMC fits to the data with the HOD model as per Section \ref{sec:MCMC_description}, with four variations. The four scenarios considered were:

A. $M_{min}$, $M_1$ and DC free (e.g. 1-halo and 2-halo amplitudes and number counts all free)

B.  $M_{1}$ fixed as function of $M_{min}$ as per Section \ref{sec:HOD_def} (e.g. 1-halo amplitude as fixed function of 2-halo amplitude)

C.  $M_{1}$ fixed as function of $M_{min}$,  $\textrm{DC}=0.6$ (e.g. 1-halo amplitude as fixed function of 2-halo amplitude, duty cycle fixed at the \citealp{Harikane2015} value)

D.  $M_{1}$ fixed as function of $M_{min}$,  $\textrm{DC}=1$ (e.g. 1-halo amplitude as fixed function of 2-halo amplitude, no duty cycle - galaxies `on' at all times)

\vspace{3mm}
In addition to these four models we also compute:
\vspace{2mm}

E. Galaxy bias from a pure bias model e.g. fit for $b$ where $\xi_{\textrm{model}}=b^{2} \xi_{\textrm{DM}}$

F. Halo masses for the most straightforwards abundance matching scheme e.g. $M_{min}$ for a sample of galaxies above a given luminosity threshold is the halo mass such that the comoving number density of halos greater than that mass is equal to the comoving number density of the galaxy sample.

\begin{figure*} 
\includegraphics[scale=0.8]{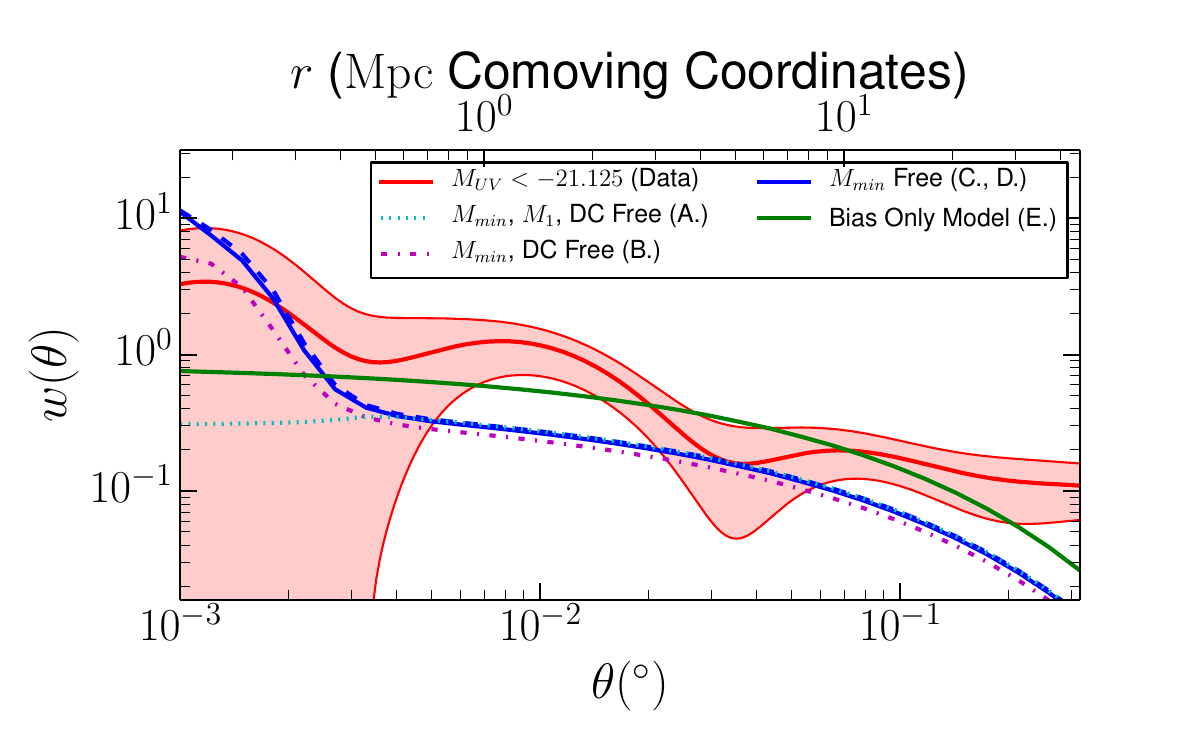}
\caption{Comparison of our measurements (red curve and shaded area) with the five different clustering models we fit. The two blue curves correspond to models with only $M_{min}$ free, the full line has $DC=0.6$ (C.) and the dashed curve $DC=1$ (D.). On linear scales all models are very similar, apart from the bias only model (E.), as the number density constraint is restricting the model from going too high. Only the $M_1$ free model (A.) allows the small-scale amplitude to vary independently.}
\label{fig:MCMC_fit_various_models}
\end{figure*}

\begin{figure*}
\includegraphics[scale=0.8]{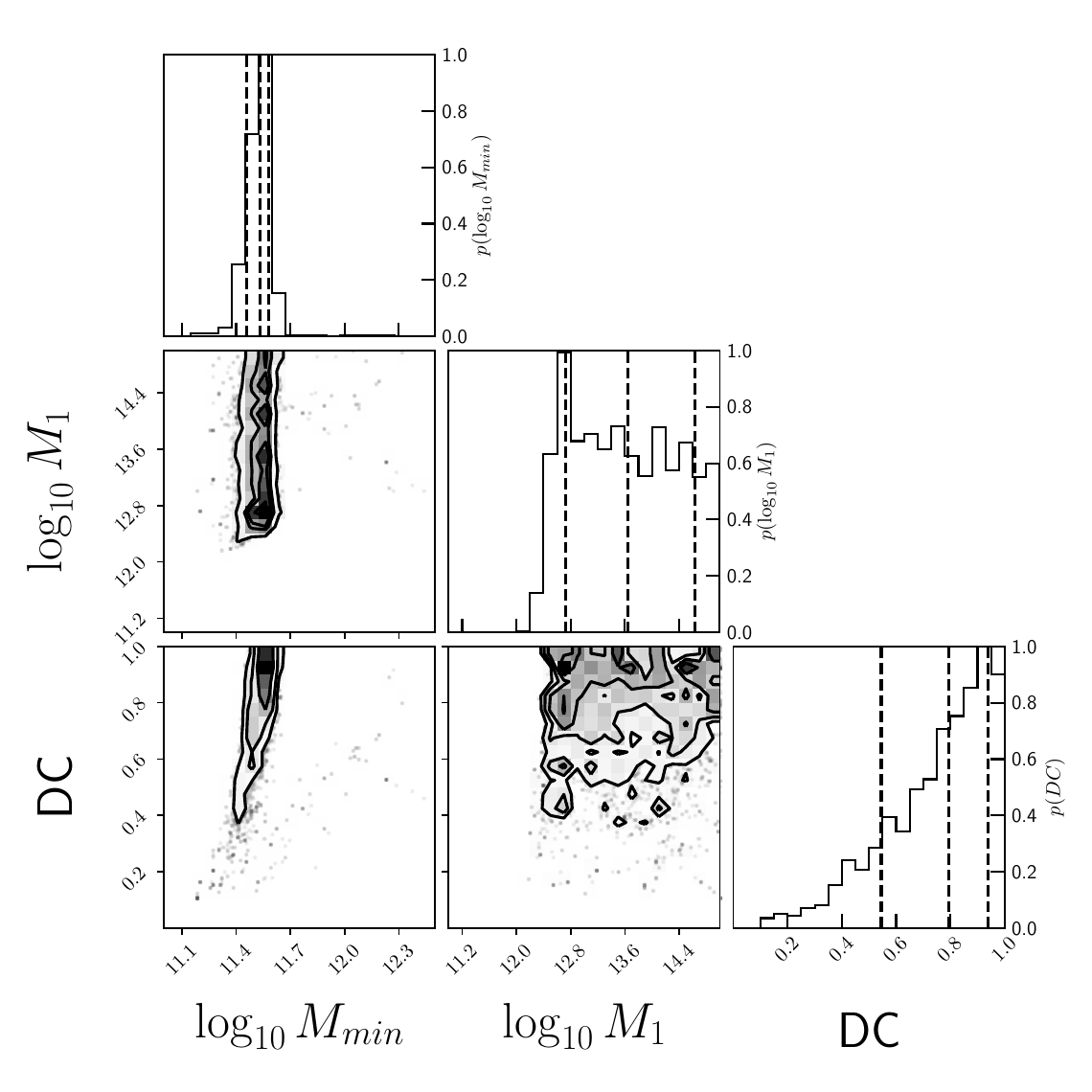}
\caption{Triangle plot of our posterior from our MCMC fitting for the HOD model with $M_{min}$, $M_{1}$ and DC free (masses in log base ten Solar mass units). The top sub-figure in each column shows the 1D marginalised posteriors (probability density functions scaled to have a peak of unity) of each parameter, with vertical dashed lines showing the 16th, 50th and 84th percentiles. The other three sub-plots show each possible 2D marginalised posterior.}
\label{fig:posterior}
\end{figure*}

The results from these 6 models are shown in Table \ref{table:summary_of_models}. Fig. \ref{fig:MCMC_fit_various_models} shows the data and the best-fit models. We show the posterior from one fitting in Fig. \ref{fig:posterior} for illustrative purposes. It is clear that the amplitude of the correlation function is roughly two orders of magnitude larger than the dark matter correlation function in the linear regime, corresponding to a very high bias. Most of our models suggest that $M_{min} \sim 10^{11.5} M_{\sun}$ e.g. our galaxies are hosted by halos of that mass and above. It also seems that the satellite fraction is at most a few percent, which suggests that at most 5-6 galaxies in our sample are satellites (in the scenario that these sources were the same underlying population as the lower luminosity LBGs, the satellite fraction could have been higher as a non-trivial portion would have been from halos hosting multiple galaxies). Although models with a range of duty cycles are consistent with the data, the observations seem to favour $DC=1$ for Model A., the model with the most freedom.

In general fits were of acceptable quality (see the $\chi^2$ values in Table \ref{table:summary_of_models}). We discuss some mild tensions in the data more in section \ref{sec:compare}, but summarise the results of each model here. The $M_{min}$, $M_{1}$ and DC free model (A.) is free to go to high masses until tension between model and measured number density stop it from going higher. This model can also take $M_1$ extremely high, to bring the amplitude of the small scale clustering down to match the data. Models that do not have $M_1$ free (B., C., D.) cannot vary their small scale behaviour freely. This forces their halo masses down, as the small scale behaviour grows rapidly with $M_{min}$; if they went higher the disagreement on small scales would become much larger. When DC is free (and $M_1$ is fixed as a function of $M_{min}$, B.), the model actually prefers to go even lower than the abundance matching halo mass, and uses the duty cycle to reach agreement with the number counts. However when these models have the duty cycle fixed (C., D.), they cannot do this, so the trade off between agreeing with small scale clustering and the number counts sets the halo mass. For $DC=0.6$ (C.) to agree with the observed number counts, the intrinsic number counts must be higher than for $DC=1$ (D.), forcing the model to prefer slightly lower halo masses. The `pure bias' model (E.) was able to fit the clustering data  well, but with a higher bias, and much larger uncertainty. This illustrates that the clustering is a comparatively weak constraint in models A.-D., and that much of the constraint on the model is coming from the number counts. The reason that models A.-D. are able to achieve stronger constraints on the bias than Model E is that the number counts are (implicitly) being mapped onto a halo mass via abundance matching, and then onto a bias by the halo bias model, as opposed to the bias being measured to that precision directly (this can only happen because the satellite fraction is so low; at lower redshift number counts don't map onto bias as cleanly as the satellite fraction is in general non-trivial). Essentially the number counts set a maximum possible bias, and the clustering gives a weak constraint pulling the estimate up to that limit. There is a mild mismatch between the clustering and the number counts - halos of the halo mass implied by the (directly measured) bias are far rarer than the observed galaxies are\footnote{This problem would have been even worse if the HMF had been taken from \cite{Tinker2010} without the high redshift correction of \cite{Behroozi2012}.}. This is in principle problematic\footnote{This is the opposite problem to what the duty cycle is invoked to solve - duty cycles in clustering studies of LBGs solved the issue of number counts being lower than implied by clustering, the problem here is the number counts are higher.}, but the issue has low statistical significance in the case discussed here. In addition, the fact that no 1-halo term emerges is slightly anomalous. We note that \citet{Harikane2015} use fitting formulae of the HMF in \citet{Tinker2010} directly without the normalization constraint, which overestimates the abundance of the HMF by a factor of ~1.7 at $z=4$ (Y. Harikane 2017, private communication, see also \citealp{Harikane2017}). So the results of \citet{Harikane2015} are likely more consistent with a duty cycle of 1 (rather than the fixed value of 0.6 that they use in their analysis). Table \ref{table:summary_of_abundance_matching} shows the observed comoving number densities, compared with the comoving number densities implied by their HOD models (alongside corresponding figures from this work).

\begin{table*}
\caption {Our constraints on the HOD parameters from the MCMC fitting. Also shown are the corresponding satellite fractions ($f_{\textrm{sat}}$) and galaxy biases ($b$) and fit reduced $\chi^2$ of the samples. Quantities in brackets are either fixed in the model, or fixed as a function of other parameters in the model. Masses are in Solar mass units (log base ten). Note that values and error bars quoted are the 16th, 50th and 84th percentiles of the posterior, as opposed to the peak values. This makes very little difference apart from the posterior for the duty cycle value for the $M_{min}$, $M_1$ and DC free model, which is peaked at DC=1 and hence only has one tail, see Fig. \ref{fig:posterior}. The lower luminosity parameter values are taken directly from \citet{Harikane2015}, apart from satellite fraction, which we calculate.}
\begin{tabular}{ ||p{2.5cm}|p{1cm}|p{1cm}|p{1.5cm}|p{1.5cm}|p{0.5cm}|p{0.5cm}|p{1cm}|p{1cm}|p{1cm}|p{1cm}|  }
\hline
Model & $M_{UV}$ & $\log M_{min}$ &  $\log M_{1}$ & $\log M_{0}$ & $\alpha$ & $\sigma$  & DC & $10^{2}f_{sat}$ & $b$ & $\chi^{2}/\textrm{d.o.f.}$ \\
\hline
A. $M_1$, DC free &-21.125 & {$11.53\substack{+0.05\\ -0.07}$}	& {$13.64\substack{+0.99 \\ -0.91}$} &	{($12.67\substack{+0.75 \\ -0.69}$)}	&	(1)	&	(0.2)	&	{$0.79\substack{+0.15 \\ -0.25}$}	&	{$<0.2$}	&	{$8.28\substack{+0.23 \\ -0.32}$}	&	1.3\\
B. DC free &-21.125 & {$11.35\substack{+0.13\\ -0.05}$}	& {($12.12\substack{+0.16 \\ -0.05)}$} &	{($11.51\substack{+0.12 \\ -0.04}$)}	&	(1)	&	(0.2)	&	{$0.36\substack{+0.3 \\ -0.12}$}	&	{$3.87\substack{+0.24\\ -0.64}$}	&	{$7.65\substack{+0.57 \\ -0.18}$}	&	1.5\\
C. DC $=0.6$ &-21.125 & {$11.48\substack{+0.02\\ -0.02}$}	& {($12.26\substack{+0.03 \\ -0.03)}$} &	{($11.62\substack{+0.02 \\ -0.02}$)}	&	(1)	&	(0.2)	&	{$(0.6)$}	&	{$3.29\substack{+0.11\\ -0.1}$}	&	{$8.16\substack{+0.1 \\ -0.1}$}	&	1.3\\
D. DC $=1$ &-21.125 & {$11.51\substack{+0.02\\ -0.02}$}	& {($12.3\substack{+0.03 \\ -0.02)}$} &	{($11.65\substack{+0.02 \\ -0.02}$)}	&	(1)	&	(0.2)	&	{$(1)$}	&	{$3.16\substack{+0.08\\ -0.1}$}	&	{$8.3\substack{+0.12 \\ -0.08}$}	&	1.9\\
E. Bias Only &-21.125 & {NA}	& {NA} &	{NA}	&	{NA}	&	{NA}	&	{NA}	&	{NA}	&	{$11.4_{-6.7}^{+6.5}$}	&	1.0\\
F. Matching Only &-21.125 & {11.51}	& {NA} &	{NA}	&	{NA}	&	{(0)}	&	{(1)}	&	{(0)}	&	{NA}	&	NA\\
\hline
Harikane16 	&	-20.0 		& {$11.30\substack{+0.10 \\ -0.13}$}	&	{($12.06\substack{+0.07 \\ -0.16})$}	&	{($11.47\substack{+0.05 \\ -0.12}$)}	&	(1)	&	(0.2)	&	(0.6)		&	5.0	&	{$6.3\substack{+0.4 \\ -0.4}$}	&	0.5\\
Harikane16 	&	-19.1 	& {$11.03\substack{+0.05 \\ -0.18}$}	&	{($11.75\substack{+0.20 \\ -0.29})$}	&	{($11.23\substack{+0.15 \\ -0.22}$)}	&	(1)	&	(0.2)	&	(0.6)		&	7.1	&	{$5.5\substack{+0.2\\ -0.4}$}	&	1.4\\
\hline
\end{tabular}
\label{table:summary_of_models}
\end{table*}

\begin{table*}
\caption {Comparison of abundance matching results to clustering fits for our data. Columns are (1) LBG sample used, (2) LBG threshold absolute magnitude, (3), observed comoving number density (Mpc$^{-3}$), (4) the minimum halo mass in the most straightforwards abundance matching scheme (log base ten Solar mass units), (5) the model comoving number density (Mpc$^{-3}$) of the best fit model  HOD in this work (A. shown) and Harikane et al., (2016) without incorporating duty cycle, (6) the corresponding minimum halo mass from the HOD model (log base ten Solar mass units).}
\begin{tabular}{ ||p{2cm}|p{1cm}|p{2cm}|p{2cm}|p{2cm}|p{2cm}|  }
\hline
Data 		& 	$M_{UV}$	& $n_{g}^{\textrm{observed}}$ (Mpc$^{-3}$)	&	$\log M_{min}^{\textrm{matched}}$ 		&	$n_{g}^{\textrm{model}}$ (Mpc$^{-3}$)	&	$\log M_{min}^{\textrm{model}}$  	\\
\hline
Bowler15 		&	-21.125		& $4.1 \times 10^{-5}$ 		&	11.51						&	$6.8 \times 10^{-5}$	&	{$11.53\substack{+0.05\\ -0.07}$}	\\
Harikane16 	&	-20.0	 	& $3.8 \times 10^{-4}$		&	11.09						&	$2.1 \times 10^{-4}$		&	{$11.30\substack{+0.10 \\ -0.13}$}	\\
Harikane16 	&	-19.1	 	& $13.4 \times 10^{-4}$		&	10.79						&	$7.3 \times 10^{-4}$		&	{$11.03\substack{+0.05 \\ -0.18}$}	\\
\hline
\end{tabular}
\label{table:summary_of_abundance_matching}
\end{table*}

\section{Discussion}

\subsection{The link between low- and high-luminosity galaxies and their haloes at $z \sim 6$} \label{sec:compare}

\begin{figure} 
\includegraphics[scale=0.45]{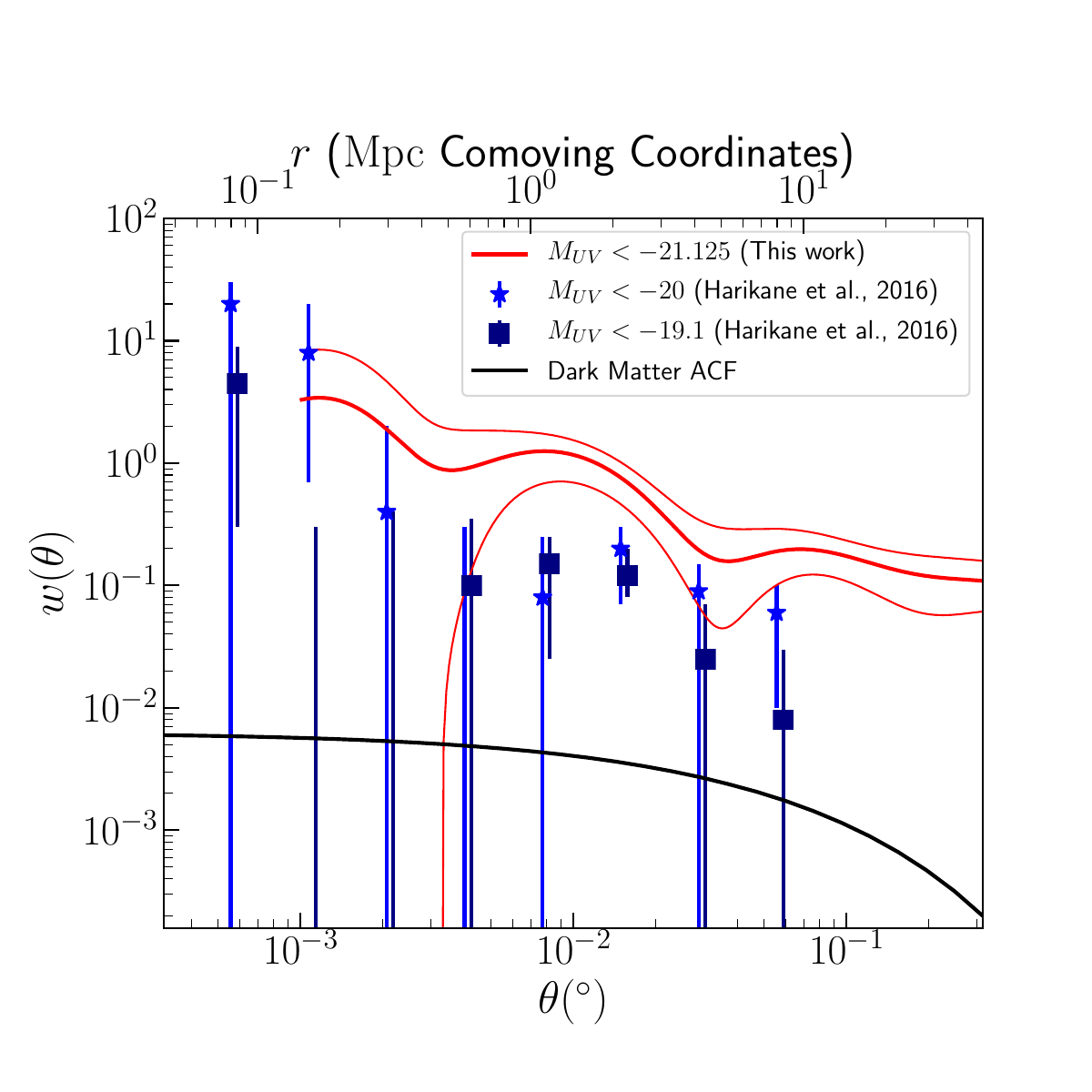}
\caption{Comparison of our measurements (red curve, 1-$\sigma$ uncertainties in the lighter curves), the lower luminosity Harikane et al., (2016) measurements, and the dark matter angular correlation function (black curve).}
\label{fig:MCMC_fit}
\end{figure}

The most relevant previous study to compare our results to is \citet{Harikane2015}, which presented clustering measurements and HOD fits to $z\sim6$ LBG galaxies, but on smaller angular scales of approximately $10^{-3.25 } <\theta / \textrm{deg} <10^{-1.25} $, compared with our $10^{-3 } <\theta / \textrm{deg}<10^{-0.5} $, and for fainter rest frame absolute magnitudes of $ -20.5<M_{UV}<-19$ compared with our $ -22.7<M_{UV}<-21.125$ sample (see Fig. \ref{fig:MCMC_fit}). Thus our results combined with \citet{Harikane2015} describe LBG clustering over almost three orders of angular scale and a factor of 40 in luminosity.

\begin{figure} 
\includegraphics[scale=0.45]{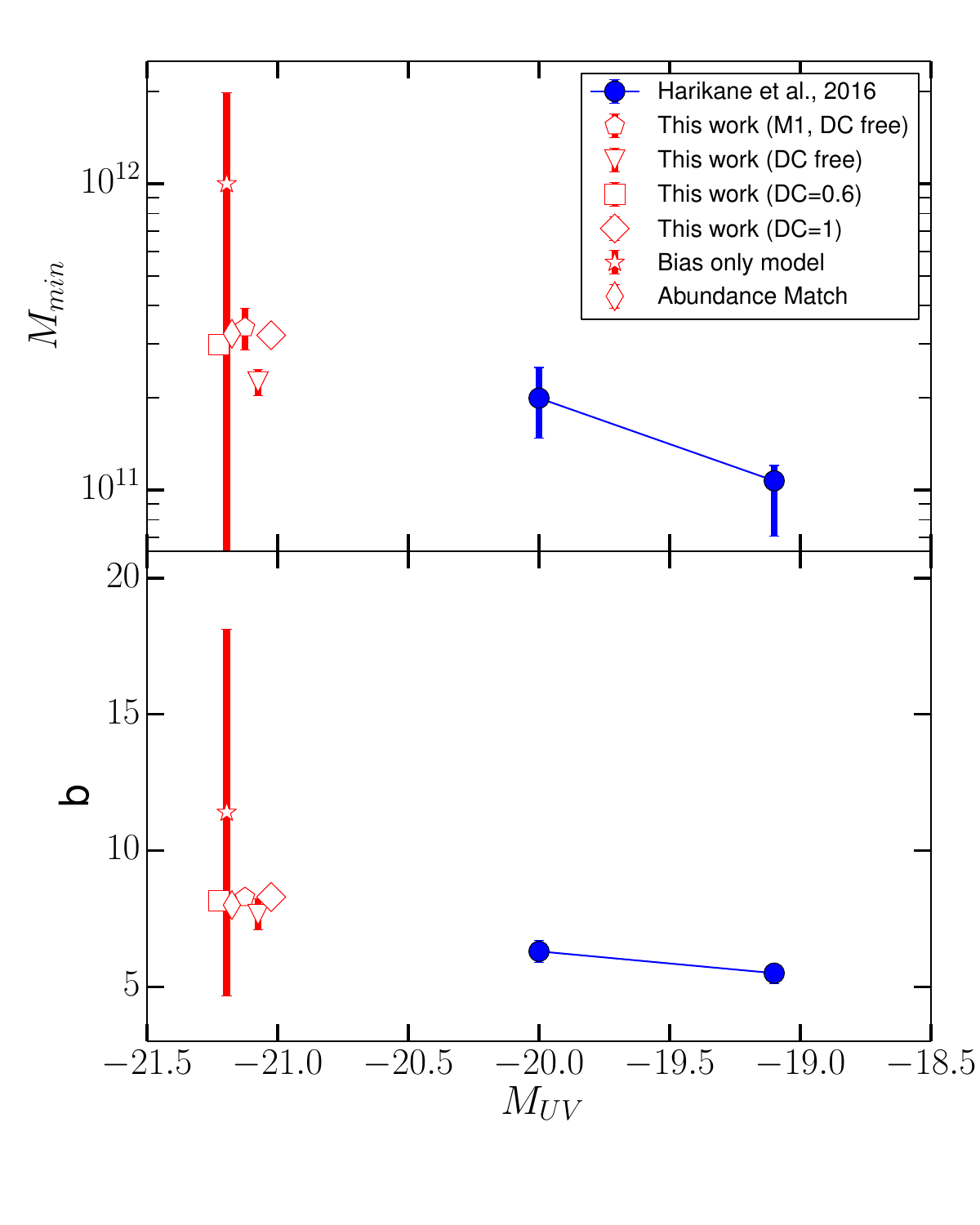}
\caption{Comparison of our results with comparable measurements of lower luminosity LBGs from \citet{Harikane2015}. Top plot: $M_{min}$ as a function of absolute UV luminosity threshold (in units of Solar mass). Bottom plot: galaxy bias as a function of absolute UV luminosity threshold. The results from our six different models are shown for comparison (x-axis values slightly offset for each model for clarity).}
\label{fig:harikane_compare}
\end{figure}

Our bias and halo mass results compared with the results of \citet{Harikane2015} are shown in Fig. \ref{fig:harikane_compare}. Although only moderate quality fits (possible reasons for which are discussed in the subsequent sub-sections), all our fitted models suggest our galaxy sample has a substantially higher typical host halo mass and galaxy bias than the lower luminosity samples in \citet{Harikane2015}. This higher bias is evident by directly comparing the two measurements of the correlation function. Our sample has an amplitude $\sim 3$ times higher than the \citet{Harikane2015} bright ($M_{UV}<-20$) sample with $\omega(0.01^{\circ}) \sim 0.2$, and our measured bias is a factor of 1.7 greater than that measured for the lower luminosity \citet{Harikane2015} sample (as $\omega \propto b^2$). In general higher luminosity and higher stellar mass galaxy samples have higher biases, but it is important to note that it was not a foregone conclusion to measure a bias this high. It was entirely possible that our $M_{UV} \sim -21.5$ sample could have been the same (or largely the same) population as the sample of \citet{Harikane2015}, just observed during a particularly vigorous but rare burst of star formation. If that had been the case, we would have measured a lower clustering amplitude, and inferred a much lower duty cycle. The comoving space density of the galaxies in our sample is $4.1 \times 10^{-5} \textrm{ Mpc}^{-3}$,  compared with $3.8 \times 10^{-4} \textrm{ Mpc}^{-3}$ for the most luminous $z=6$ \citet{Harikane2015} sample. \citet{Harikane2015} do not measure the duty cycle for this sample, but assume it to be equal to 0.6. As an illustrative example, a duty cycle of 0.6 for the \citet{Harikane2015} sample would mean an actual underlying population comoving density of $6.3\times 10^{-4} \textrm{Mpc}^{-3}$. If our sample was part of the same population, that would correspond to $DC=0.06$ (in other words that the fainter population spends approximately 6 percent of its time in this super-enhanced state of star formation). However the amplitude of the clustering rules this out and our $M_{UV} < -21.125$ sample is comprised of continuously high-luminosity objects in very dense environments. Although the measurement of the correlation function for the UDS field is too poor to make a measurement of the bias, the bias (Model E) from just the UltraVISTA field is 
$b=11.1\substack{+7.6 \\ -6.8}$, consistent with this result. In future larger data sets it will be important to perform the full HOD analysis for separate fields individually to understand how cosmic variance impacts inferred HOD parameters. Planned work with $z\sim6-7$ LBG samples selected in the VIDEO survey (which goes to comparable depths to the analysis in this work) will allow this comparison to be done with 12deg$^2$ of data over three roughly equally sized independent fields.

\cite{Bowler2014} and \cite{Bowler2015} report a rapid evolution in the high luminosity end of the luminosity function at $z=6-7$, a transition from a power-law drop off to an exponential cut-off, interpreted as the onset of quenching or dust obscuration. Our clustering results show that these galaxies at the bright end of the luminosity function truly are in the densest regions of the Universe, as opposed to less biased objects caught in extremely rare massive star-bursts. This result gives extra support to the claim that the high-redshift evolution of the high-luminosity end of the luminosity function is determined by dust extinction, cooling rate of gas in halos, or feedback processes, as opposed to how rare its episodes of high star formation are.

Preliminary results from the Great Optically Luminous Dropout Research Using Subaru HSC LBG survey (GOLDRUSH, \citealp{Ono2017}) have started to give complementary results over an even larger volume, with deep optical imaging over 100 deg$^2$. \citet{Ono2017} suggests that at these redshifts the high luminosity end of the luminosity function is indeed best fit by a double power law, supporting \citet{Bowler2014} and \citet{Bowler2015}. Similarly, \citet{Harikane2017} make clustering measurements in GOLDRUSH that suggest clustering amplitude continues to increase at high luminosities, and that the ratio of luminosity to halo mass remains high at high halo masses at $z\sim6$. However their results currently only use optical imaging (unlike our sources, which were selected using near-infrared bands), which makes processes like removing brown dwarf contamination more complex.

\subsection{Apparent lack of a 1-Halo term} \label{sec:no_1_halo}

Models using extrapolated values of $M_1$ (B., C., D.) suggest that at scales of $10^{-2.5}$deg and smaller there should be a sharp upturn in the value of the correlation function as the observations start to probe clustering of multiple galaxies within individual halos (see Fig. \ref{fig:MCMC_fit_various_models}). We do not observe this in the data, in contrast to \cite{Harikane2015}, see Fig. \ref{fig:MCMC_fit}. A direct interpretation of this would be that $M_1$ just increases much faster than the extrapolation of Equation \ref{eq:M1} i.e. the satellite fraction drops off extremely fast and an unfeasibly large (for this redshift) halo is needed to host two of these sources. Another possibility is suggested by \cite{Jose2013}, who also observe a lack of a 1-halo term in clustering measurements of $z \sim 3-7$ LAEs. Their proposed solution was that halo occupancy behaved in a sub-Poissonian manner, and they found that a modified distribution (see their equation 15) was able to reproduce the measurements.  However, we suggest that there are good reasons to believe that there is strong cosmic variance on our small scale measurements that is not accounted for in the bootstrap uncertainties, making it hard to make direct inferences about the satellite population of these galaxies.

For single contiguous field observation, cosmic variance is smaller on small scales than on large scales in a limited sense, simply because one observes more instances of small scale structure. However these are not independent instances of small-scale structure, as they all come from the same large-scale density field. To illustrate this, suppose our sources have $M_{\textrm{min}} \sim10^{11.4} M_{\sun}$, then we would expect $M_{1} \sim 10^{12.2} M_{\sun}$ e.g. only halos with $M>10^{12.2} M_{\sun}$ host more than one of our bright sample. The comoving density of $M>10^{12.2} M_{\sun}$ halos is $5.2 \times 10^{-7} \textrm{Mpc}^{-3}$ and the comoving volume probed by the observations is $\sim 1.7 \times 10^{7} \textrm{Mpc}^{3}$.  This means that the expected numbers of $M>10^{12.2} M_{\sun}$ halos in the volume surveyed is $\sim 10$. Just these ten would give $\sim 10$ close pairs (in addition a few more would be expected from projection effects), which is more than twice the 4 close ($10^{-3} - 10^{-2.5}$deg) pairs observed here, and would push the small scale correlation function up. However these halos will be extremely biased, much more than the $ \sim10^{11.4} M_{\sun}$ halos. Conceivably for an extreme case, it could be that if our observations were repeated 10 times, we would find that 9 times no $M>10^{12.2} M_{\sun}$ halos were observed, and the tenth time a very overdense region is observed, which has 100 in. In the first nine cases, no satellites would be observed, leading to a flat correlation function to small radii that we see in our observations, and the tenth time an overestimate of the satellite fraction is measured. Therefore we would expect there to be very substantial cosmic variance on our measurements of the correlation function on small scales - variance that is not incorporated into our errors on our clustering measurements. Essentially the 1-halo term is dominated by contributions from very massive halos, which are the most biased, so there is the most cosmic variance on small scale measurements of the correlation function. This is also consistent with our less dense field (UDS), having no power on small scales. It appears to be the case that neither of our fields are overdense enough to sample the highly biased sample of massive halos at this redshift which could be massive enough to host multiple bright LBGs, and instead our small scale measurements are dominated by the angular projection of the linear clustering (e.g. objects near in angular space by chance, but not near in physical space).

The \cite{Harikane2015} measurements however do have a prominent 1-halo term. We suggest that the reason for this may lie in the fact that a) they are at lower luminosities, so the cosmic variance on halos required to host multiple galaxies is less extreme and b) their correlation functions are measured from galaxies in seven different fields (rather than our two), so they had a greater chance of observing a dense field that had the massive halos necessary for satellites.

\subsection{Mild mismatch between the number counts and bias measurements}

As discussed in Section \ref{sec:full_sample_modelling}, the bias obtained by fitting just to the clustering (Model E) is higher than that obtained when fitting to both the counts and the clustering. The core of the discrepancy is that to obtain the directly measured bias of  $\sim 11$,  the galaxies would need to be in halos of minimum mass $M_{\textrm{min}} \sim 10^{12} M_{\sun}$. This corresponds to a comoving density of $\sim 2 \times 10^{-6} \textrm{Mpc}^{-3}$, compared with an observed number density of $4.1 \times 10^{-5} \textrm{Mpc}^{-3}$ i.e. approximately a factor of 20 lower than the observed value. Equivalently, plain abundance matching (Model F.) would suggest a minimum mass $M_{\textrm{min}} \sim 10^{11.5} M_{\sun}$ (see table \ref{table:summary_of_abundance_matching}) corresponding to a bias of $\sim 8$. We note that \citet{Barone-Nugent2014} report a very similar issue at $z \sim 7.2$, where they found that a duty cycle of 1 was needed for their LBG sample, and that even then the measured bias was slightly inconsistent with the number density.

Although of comparatively low statistical significance, we consider some possible explanations for this discrepancy:

\begin{itemize}
\item Contaminants: some of the sources used for our clustering measurements are not truly $z \sim 6$ LBGs, but are instead brown dwarfs or galaxies at other redshifts. This is  unlikely to be the cause of the discrepancy, as \cite{Bowler2015} had access to photometry across a very large range of wavelengths and performed extensive testing with brown dwarf templates to rule out substantial contamination. Furthermore, stellar contamination would actually reduce the clustering amplitude as stars are unclustered and have no physical correlation with the galaxies. In addition, \citet{Bowler2014} and \citet{Bowler2015} investigated the possibility of contamination from lower luminosity sources being magnified by gravitational lensing. They found that lensing was not a significant factor in either the selection of the sources, nor the constraints on the luminosity function, based on the non-detection of strong lensing sources along the lines of sight, and the fact that the sources have no excess lensing over random locations in the field.
\item More complex galaxy-halo relations: halo occupancy depends on more than just halo mass. HOD modelling is predicated on the principle that the only thing that determines the galaxy content of a halo is the mass of the halo - if this is violated, then in general more complex relations between the galaxy-halo relation and clustering measurements are possible. Well known cases include assembly bias (see \citealp{Hearin2015}, where the bias of halos depends on halo assembly history as well as mass), or a dependence on the large scale linear density field. Galaxy-galaxy lensing can in principle observationally break these degeneracies (e.g. if sources are in older, lower mass halos, their clustering will reflect the assembly-dependent bias of the  hosts, but the lensing will reflect just the mass), but this is likely to never be possible at these redshifts as it requires a high number density of even higher redshift sources. It seems likely that comparison with simulations is the only way to investigate the viability of such underlying processes.
\item Uncertainty in knowledge of the high-redshift dark matter distribution: that the halo mass function etc. used in the HOD model are flawed. \cite{Tinker2008} and \cite{Tinker2010} found model HMFs and halo biases from N-body simulations at redshifts of $z=0-2.5$. \cite{Behroozi2012} then introduced a high-redshift calibration to the \cite{Tinker2010} HMF, extending the validity to $z \sim 8$ (representing an increase of approximately 20 percent at $z=6$ for $M \sim10^{11.3} M_{\sun}$ halos). However, \cite{Behroozi2012} did not calibrate the high redshift halo bias, so we are effectively using biases at $z \sim 2.5$ extrapolated to $z \sim 6$. The excess in the clustering amplitude is only around a factor of 50 percent, which would require around a 25 percent correction in bias. Thus we suggest that our results could potentially be explained with a high-redshift calibration to the halo bias function that steepens it at the high mass end. See also \cite{Behroozi2016} for a discussion on this direction of inference e.g. how high redshift stellar-mass functions can give information on the high-redshift HMF. An alternate potential correction to our understanding of the distribution of dark matter is the  incorporation of `quasi-linear effects'. HOD modelling makes a binary division between non-linear clustering within halos, and large-scale linear bias. However this transition is gradual, not sharp, and bias can be scale-dependent on up to 10 Mpc scales  (relevant for scales probed with our observations), although it always tends to a constant value at large scales (\citealp{Mann1997a}). Introducing a functional form for scale-dependent bias can model some of these effects and \cite{Jose2016} conclude that quasi-linear clustering has the largest effect at high redshift ($z>2$), and high halo mass. In particular, \cite{Jose2017} note a similar discrepancy to ours at $3<z<5$, and find that quasi-linear effects can cause one to over estimate halo mass by up to a factor of ten if unaccounted for. 
\end{itemize}

In summary there are a few possible ways for the measured bias to be higher than the number counts would seem to permit. Significant contamination seems unlikely, and uncertainty in the modelling of the distribution of dark matter in the high redshift Universe would seem a simpler issue to consider before invoking more complex galaxy-halo connections. Although the discrepancy has a low statistical significance, it seems that incorporation of a high-redshift halo bias calibration, or quasi-linear bias could be worthwhile in future analyses.

\subsection{Estimating Cosmic Variance} \label{sec:cosmic_variance}

Cosmic variance is a term that can be used to refer to a number of related but subtly different effects. The specific context in which we use the term here is that many extragalactic statistical measurements vary by more than sample variance between different fields because of large scale structure. As noted in \cite{Bowler2015}, the number density of our two fields varies by much more than sample variance assuming a Poisson distribution. This is a consequence of large-scale structure, which our clustering measurements quantify. These clustering measurements can be linked back to the number count estimates to see if the cosmic variance observed is consistent with the clustering measurements, or if one of the fields is over/under dense, even accounting for large-scale structure. Understanding cosmic variance can be important for correctly connecting high redshift observations of galaxies with our understanding of reionization e.g. \cite{Ouchi2009}.

Note that in general it is possible for two populations to have the same average number counts, but different cosmic variances - this occurs when they have the same 1-point statistics, but different 2-point statistics. Thus we can use the 2-point statistics to refine the estimate of cosmic variance in \citet{Bowler2015} who used the Trenti Cosmic Variance calculator (\citealp{Trenti2008}) - which only uses 1-point statistics.  A clustered and unclustered population of the same number density will have substantial and zero cosmic variance respectively (both will have Poisson variance).

The cosmic variance is related to the \textit{expected value} of the correlation function in the geometry of the field, that is to say the expected value of the correlation function at the separation of two points randomly selected in the field. Analytically we can write the expectation as:

\begin{equation}
\bar{\omega}(A)=\frac{\int_{A}\int_{A}\omega(|\vec{\theta_{i}}-\vec{\theta{j}}|)d^{2}\theta_{i} d^{2}\theta_{j}}{\int_{A}\int_{A}d^{2}\theta_{i} d^{2}\theta_{j}} , 
\end{equation}

where $A$ is the angular region of the field, $\omega$ is the 2-point correlation function, $\bar{\omega}(A)$ is the expectation of the correlation function in that field, $\theta_{i}$ and $\theta_{j}$ are points in the field, $|\vec{\theta_{i}}-\vec{\theta{j}}|$ is their angular separation, and the integrals are double integrals over the area of the field. We calculate this numerically by sampling 100,000 pairs of points in the field, calculating their angular separation, finding the value of the correlation function at that angular scale (with the best fit model from Section \ref{sec:full_sample_modelling}), and then taking the average.

We use the formalism of \citet{Trenti2008} that summarises well established results from \citet{Peebles1980}, \citet{Peebles1993}, \citet{Newman2001} and \citet{Somerville2003} that conclude:

\begin{equation}
\bar{\omega}(A)=\frac{\langle N^{2} \rangle - \langle N \rangle^2}{\langle N \rangle^2} - \frac{1}{\langle N \rangle}
\end{equation}

where $N$ is the random variable of the number of objects in a field. This can be rearranged to the form:

\begin{equation}
\langle N^{2} \rangle - \langle N \rangle^2=\langle N \rangle+\bar{\omega}(A) \langle N \rangle^{2} ,
\end{equation}

\begin{equation*}
\sigma^{2}_{total}= \sigma^{2}_{poisson}+\sigma^{2}_{CV} ,
\end{equation*}

where $\sigma_{total}=\sqrt{\langle N^{2} \rangle - \langle N \rangle^2}$ is the total standard deviation on measurements of number counts, $\sigma_{Poisson}=\sqrt{\langle N \rangle}$ is the Poisson standard deviation and $\sigma_{CV}=\sqrt{\bar{\omega}(A)} \langle  N \rangle$ is the standard deviation from cosmic variance e.g. total standard deviation is the Poisson and cosmic variance standard deviations added in quadrature. The standard deviation from cosmic variance reduces to $\sigma_{CV}=b \sqrt{\bar{\omega}_{DM}(A)} \langle  N \rangle$ in the `pure-bias' case where $\omega=b^2\omega_{DM}$, where $b$ is the bias and $\omega_{DM}$ is the dark matter angular correlation function.

This formalism has all the properties one would expect from cosmic variance. Cosmic variance is higher when sources are more clustered (further away from uniform). Cosmic variance becomes lower as the size of the field increases, as the correlation function is sampling larger scales, where the function has a lower value, a consequence of the fact that a larger range of environments are being probed. A more subtle effect is that cosmic variance also varies with field shape, as well as size. The average length scale probed for a circle is a lot smaller than for a long thin rectangle of the same area (for example), corresponding to a higher average correlation function value, and greater cosmic variance. This can be interpreted (as described in \citealp{Trenti2008}) as a consequence of the fact that a more compact field geometry is predominantly sampling the same environment, be it an over- or under-density. However a long thin geometry is sampling from a large range of environments, and overdensities and underdensities are more likely to cancel out. The formalism for describing cosmic variance here also works when the field is disconnected. If the `field' is actually two disconnected subfields separated by a vast distance in the sky, when calculating the average of the correlation function over this field, half the time the two points will be in different sub-fields, and the value of the correlation function on this scale will be effectively zero. This halves the value of $\bar{\omega}(A)$, effectively reducing cosmic variance contribution by a factor of $\sqrt{2}$ as completely different regions of the Universe are being probed.

\begin{table*}
\caption {Actual and expected number of galaxies in each field. The columns are: field used, the field angular area (in deg$^2$),  the actual number of galaxies in the field ($N_{\textrm{a}} $),  the actual angular galaxy density in the field ($\rho_{\textrm{a}}$, in deg$^{-2}$), the expected number of galaxies in the field if it had the mean density based on the two fields combined ($N_{\textrm{e}} $), the expected angular galaxy density in the field - all identical figures as we are considering deviations from the mean density ($\rho_{\textrm{e}}$, in deg$^{-2}$), the expected value of the correlation function over the field ($\bar{\omega}$), standard deviation of the surface density estimate from Poisson statistics based on the square root of $N_{\textrm{e}} $ ($\sigma_{Poisson}$, in deg$^{-2}$), standard deviation on the surface density from cosmic variance, estimated from our clustering measurements ($\sigma_{CV}$, in deg$^{-2}$), our Poisson and cosmic variance errors added in quadrature ($\sigma_{Total}$, in deg$^{-2}$)}
\begin{tabular}{ |p{2cm}||p{1cm}|p{1cm}|p{1cm}|p{1cm}|p{1cm}|p{1cm}|p{1cm}|p{1cm}|p{1cm}|  }
 \hline
Field & Area (deg$^{2}$)  & $N_{\textrm{a}} $   & $\rho_{\textrm{a}}$ (deg$^{-2}$)   & $N_{\textrm{e}}$ & $\rho_{\textrm{e}}$ (deg$^{-2}$) &  $\bar{\omega}$	&	$\sigma_{Poisson}$ (deg$^{-2}$) &  $\sigma_{CV}$ (deg$^{-2}$) &  $\sigma_{Total}$ (deg$^{-2}$) \\
 \hline

UDS			&	0.74	&	64	&	86 	&	92	&	124	&	0.031		&	13	&	22		&	25\\
UltraVISTA	&	0.62	&	103	&	166	&	77	&	124	&	0.026		&	14	&	20		&	24\\ 
Total			&	1.35	&	167	&	124	&	167	&	124	&	0.014 	&	10	&	15		&	18\\  \hline
\end{tabular}
\label{table:summary}
\end{table*}

Table \ref{table:summary} summarises our results when applied to the UltraVISTA and UDS fields for our $z\sim6$ samples (using our best fit pure bias model, Model E). The most important columns to compare are the $\rho_{a}$ (the actual surface density of galaxies in the field), $\rho_{e}$ (the expected surface density there would be if both fields had the average density) and $\sigma_{total}$ the standard deviation on the surface density measurement including the Poisson and cosmic variance implied by our clustering measurements. The observed over/under-densities of galaxies in the UltraVISTA and UDS fields are 1.75-$\sigma$ and 1.5-$\sigma$ deviations from the model value respectively. For comparison, a straightforwards estimation of  $\sigma_{Total}$ using the conventional expression for the standard deviation would suggest $\sigma_{Total}\approx57$deg$^{-2}$ for the individual sub-fields, substantially higher than our quoted values. This suggests either that our estimates of the cosmic variance in Table \ref{table:summary} are underestimates, or that our fields are extreme over/under-densities. As noted in \citealp{Bowler2014}, they are the most over- and under-dense respectively of the five CANDELS fields, which lends support to the latter interpretation - the observed densities appear to be unusual/extreme over/under-densities. For reference, the comoving number density of Model A (see table \ref{table:summary_of_abundance_matching}) would correspond to a source surface density of  $\sim169$ deg$^{-2}$ when including duty cycle. To summarise, at these redshifts, both UltraVISTA and UDS appear to be moderate, but not unreasonable, over and under-densities respectively. Future work will seek to also better understand a) the cosmic variance uncertainty on our clustering measurements (imperfectly accounted for by our bootstrap-object approach) alongside the cosmic variance on the number counts, and b) the impact of cosmic variance on inferred HOD parameters by modelling the measurements separately in each of the VIDEO fields.

\subsubsection{Estimating the Cosmic Variance on the Clustering Measurements} \label{sec:clustering_cosmic_variance}

Our results in table \ref{table:summary} suggest that cosmic variance still contributes substantially to the uncertainty on our number counts, even though we are using two highly independent fields. This suggests the uncertainties on our clustering measurements could also be substantially underestimated, leading to underestimates of the uncertainty on inferred parameters. As discussed earlier, a full examination of the role of cosmic variance on the clustering measurements is beyond the scope of this work. However, to gain an approximate estimate of the impact of cosmic variance on the inferred parameters, we make the simple ansatz of scaling the clustering covariance matrix by the amount the counts uncertainty is increased by the presence of cosmic variance i.e. $(\sigma_{Total}/\sigma_{Poisson})^{2}\approx3.2$ (shown with dashed green lines in figure \ref{fig:full_acf}). We then repeat the analysis in section \ref{sec:RESULTS} using these `cosmic variance corrected' uncertainties. We find that for Model E (bias only), this correction alters the inferred bias from $b=11.4_{-6.7}^{+6.5}$ to $b=13.0_{-8.0}^{+8.5}$. For Model A, $\log_{10} M_{min}=11.53\substack{+0.05\\ -0.07}$ becomes $\log_{10} M_{min}=11.40\substack{+0.06\\ -0.11}$ and still favours $DC \approx 1$ and $\log_{10} M_{min}>12.2$ (low satellite fraction). The true uncertainties on our inferred parameters are likely somewhere between those in section \ref{sec:RESULTS} and those with the `corrected' uncertainties (as bootstrap-objects does not \textit{completely} neglect cosmic variance) - but these results using the corrected uncertainties illustrate that our key conclusions are likely unaltered even when cosmic variance on clustering measurements is accounted for.

\section{Conclusions} \label{sec:conclusions}

We have used the largest existing sample of extremely bright Lyman-break galaxies at $z\sim6$ to investigate their large scale structure and links to the possible onset of feedback quenching or dust obscuration at this redshift. This sample (detailed in \citealp{Bowler2015}) of 263 LBGs was selected in the UltraVISTA/COSMOS and UDS/SXDS fields, using deep optical and near-infrared data required  to distinguish the galaxies from contaminant populations. The method we used to study the connection between the galaxies and their host halo was to measure their clustering with the angular correlation function, and model these measurements with a HOD scheme.

The key conclusions of this work are:
\begin{itemize}
\item Bright LBGs ($M_{UV}\leqslant-21$) appear to be highly biased ($b \sim 11\pm7$ based on the clustering, favouring $b \sim 8$ if the number counts are also taken into account) objects in dense environments, as opposed to being rare temporal episodic incarnations of fainter galaxies ($M_{UV} \sim -19$). Our results have important implications for the physical origin of the observed steepening of the bright end of the ultra-violet luminosity function between $z \sim 6$ and $z \sim 7$ (\citealp{Bowler2014,Bowler2015}) - which in a straightforward abundance matching scheme would imply a dramatically increased luminosity to halo mass ratio at $z \sim 7$ to $z \sim 6$. This measured high bias potentially suggests that the bright-end of the luminosity function at $z \sim 6$ could be determined by feedback processes or dust obscuration, rather than duty cycles.
\item The bias based on the clustering is higher than that suggested by the number counts, potentially suggesting that some modification to our knowledge of the high-redshift dark matter distribution could be needed This is most likely to be the incorporation of quasi-linear effects (as described in \citealp{Jose2017}), or possibly a minor calibration upwards of halo bias at high redshift.
\item Although number counts within each field differ by far more than Poisson sample variance, estimates of the cosmic variance from the clustering would suggest that both fields are reasonably extreme $\sim1.6$-$\sigma$ over/under densities.
\item We do not require duty cycle to explain our observations (equivalently $DC \sim 1$), and the satellite fraction of the sources is very small, at most a few percent
\end{itemize}

In the next few years, deep, wide surveys such as VIDEO and the VISTA Extragalactic Infrared Legacy Survey (VEILS, \citealp{Honig2016}), which will extend the area of VIDEO, will provide improved constraints on the luminosity function and clustering of high-redshift galaxies, and allowing extension to even more luminous LBGs. By the mid 2020s it should be possible to use the \textit{Euclid} space telescope mission to do this with 1,000s of LBGs (\citealp{Bowler2016a}), which will reveal how the measured large scale structure of LBGs and LAEs relates to reionization.

\section*{Acknowledgements}

The authors thank the anonymous referee for the comments that have improved the quality of this paper. PH and CH wish to acknowledge support provided through STFC studentships. PH wishes to thank the Rector and Fellows of Lincoln College for support through the Graduate Research Fund. Many thanks to Yuichi Harikane and Olmo Piana for useful discussions on high redshift clustering, and to Steven Murray for advice for using {\sc Halomod}. This work was supported by the Oxford Centre for Astrophysical Surveys which is funded through generous support from the Hintze Family Charitable Foundation, the award of the STFC consolidated grant  (ST/N000919/1), and the John Fell Oxford University Press (OUP) Research Fund. Based on data products from observations made with ESO Telescopes at the La Silla or Paranal Observatories under ESO programme ID 179.A- 2006. Based on observations obtained with MegaPrime/MegaCam, a joint project of CFHT and CEA/IRFU, at the Canada-France-Hawaii Telescope (CFHT) which is operated by the National Research Council (NRC) of Canada, the Institut National des Science de l'Univers of the Centre National de la Recherche Scientifique (CNRS) of France, and the University of Hawaii. This work is based in part on data products produced at Terapix available at the Canadian Astronomy Data Centre as part of the Canada-France-Hawaii Telescope Legacy Survey, a collaborative project of NRC and CNRS.

\bibliographystyle{mn2e_mod}
\bibliography{LBG_paper}

\begin{thebibliography}{112}
\expandafter\ifx\csname natexlab\endcsname\relax\def\natexlab#1{#1}\fi

\bibitem[{Adam {et~al}\mbox{.}(2016)Adam, Aghanim, Ashdown, Aumont,
  Baccigalupi, Ballardini, Banday, Barreiro, Bartolo, Basak, Battye, Benabed,
  Bernard, Bersanelli, Bielewicz, Bock, Bonaldi, Bonavera, Bond, Borrill,
  Bouchet, Boulanger, Bucher, Burigana, Calabrese, Cardoso, Carron, Chiang,
  Colombo, Combet, Comis, Couchot, Coulais, Crill, Curto, Cuttaia, Davis,
  de~Bernardis, de~Rosa, de~Zotti, Delabrouille, {Di Valentino}, Dickinson,
  Diego, Dor{\'{e}}, Douspis, Ducout, Dupac, Elsner, En{\ss}lin, Eriksen,
  Falgarone, Fantaye, Finelli, Forastieri, Frailis, Fraisse, Franceschi,
  Frolov, Galeotta, Galli, Ganga, G{\'{e}}nova-Santos, Gerbino, Ghosh,
  Gonz{\'{a}}lez-Nuevo, G{\'{o}}rski, Gruppuso, Gudmundsson, Hansen, Helou,
  Henrot-Versill{\'{e}}, Herranz, Hivon, Huang, Ili{\'{c}}, Jaffe, Jones,
  Keih{\"{a}}nen, Keskitalo, Kisner, Knox, Krachmalnicoff, Kunz, Kurki-Suonio,
  Lagache, L{\"{a}}hteenm{\"{a}}ki, Lamarre, Langer, Lasenby, Lattanzi,
  Lawrence, {Le Jeune}, Levrier, Lewis, Liguori, Lilje, L{\'{o}}pez-Caniego,
  Ma, Mac{\'{i}}as-P{\'{e}}rez, Maggio, Mangilli, Maris, Martin,
  Mart{\'{i}}nez-Gonz{\'{a}}lez, Matarrese, Mauri, McEwen, Meinhold,
  Melchiorri, Mennella, Migliaccio, Miville-Desch{\^{e}}nes, Molinari, Moneti,
  Montier, Morgante, Moss, Naselsky, Natoli, Oxborrow, Pagano, Paoletti,
  Partridge, Patanchon, Patrizii, Perdereau, Perotto, Pettorino, Piacentini,
  Plaszczynski, Polastri, Polenta, Puget, Rachen, Racine, Reinecke,
  Remazeilles, Renzi, Rocha, Rossetti, Roudier, Rubi{\~{n}}o-Mart{\'{i}}n,
  Ruiz-Granados, Salvati, Sandri, Savelainen, Scott, Sirri, Sunyaev, Suur-Uski,
  Tauber, Tenti, Toffolatti, Tomasi, Tristram, Trombetti, Valiviita, {Van
  Tent}, Vielva, Villa, Vittorio, Wandelt, Wehus, White, Zacchei, \&
  Zonca}]{PlanckCollaboration2016}
Adam R. {et~al.}, 2016, Astronomy {\&} Astrophysics, 596, A108

\bibitem[{Arnouts {et~al}\mbox{.}(1999)Arnouts, Cristiani, Moscardini,
  Matarrese, Lucchin, Fontana, \& Giallongo}]{Arnouts1999}
Arnouts S., Cristiani S., Moscardini L., Matarrese S., Lucchin F., Fontana A.,
  Giallongo E., 1999, Monthly Notices of the Royal Astronomical Society, 310,
  540

\bibitem[{Bahcall \& Soneira(1983)}]{Bahcall1983}
Bahcall N.~A., Soneira R.~M., 1983, The Astrophysical Journal, 270, 20

\bibitem[{Barone-Nugent {et~al}\mbox{.}(2014)Barone-Nugent, Trenti, Wyithe,
  Bouwens, Oesch, Illingworth, Carollo, Su, Stiavelli, Labbe, \& van
  Dokkum}]{Barone-Nugent2014}
Barone-Nugent R.~L. {et~al.}, 2014, The Astrophysical Journal, 793, 17

\bibitem[{Bartelmann \& Schneider(2001)}]{Bartelmann1999}
Bartelmann M., Schneider P., 2001, Physics Reports, 340, 291

\bibitem[{Becker {et~al}\mbox{.}(2015)Becker, Bolton, \& Lidz}]{Becker2015}
Becker G.~D., Bolton J.~S., Lidz A., 2015, Publications of the Astronomical
  Society of Australia, 32, e045

\bibitem[{Behroozi \& Silk(2016)}]{Behroozi2016}
Behroozi P., Silk J., 2016, eprint arXiv:1609.04402

\bibitem[{Behroozi {et~al}\mbox{.}(2013)Behroozi, Wechsler, \&
  Conroy}]{Behroozi2012}
Behroozi P.~S., Wechsler R.~H., Conroy C., 2013, The Astrophysical Journal,
  770, 57

\bibitem[{Benson {et~al}\mbox{.}(2000)Benson, Cole, Frenk, Baugh, \&
  Lacey}]{Benson1999b}
Benson A.~J., Cole S., Frenk C.~S., Baugh C.~M., Lacey C.~G., 2000, Monthly
  Notices of the Royal Astronomical Society, 311, 793

\bibitem[{Berlind \& Weinberg(2002)}]{Berlind2001}
Berlind A.~A., Weinberg D.~H., 2002, The Astrophysical Journal, 575, 587

\bibitem[{Bertin \& Arnouts(1996)}]{Bertin1996}
Bertin E., Arnouts S., 1996, Astronomy and Astrophysics Supplement Series, 117,
  393

\bibitem[{Beutler {et~al}\mbox{.}(2011)Beutler, Blake, Colless, Jones,
  Staveley-Smith, Campbell, Parker, Saunders, \& Watson}]{Beutler2011}
Beutler F. {et~al.}, 2011, Monthly Notices of the Royal Astronomical Society,
  416, 3017

\bibitem[{Bhowmick {et~al}\mbox{.}(2017)Bhowmick, {Di Matteo}, Feng, \&
  Lanusse}]{Bhowmick2017}
Bhowmick A.~K., {Di Matteo} T., Feng Y., Lanusse F., 2017, eprint
  arXiv:1707.02312

\bibitem[{Bouwens {et~al}\mbox{.}(2007)Bouwens, Illingworth, Franx, \&
  Ford}]{Bouwens2007}
Bouwens R.~J., Illingworth G.~D., Franx M., Ford H., 2007, The Astrophysical
  Journal, 670, 928

\bibitem[{Bouwens {et~al}\mbox{.}(2015)Bouwens, Illingworth, Oesch, Trenti,
  Labb{\'{e}}, Bradley, Carollo, van Dokkum, Gonzalez, Holwerda, Franx,
  Spitler, Smit, \& Magee}]{Bouwens2015}
Bouwens R.~J. {et~al.}, 2015, The Astrophysical Journal, 803, 34

\bibitem[{Bouwens {et~al}\mbox{.}(2016)Bouwens, Oesch, Labb{\'{e}},
  Illingworth, Fazio, Coe, Holwerda, Smit, Stefanon, van Dokkum, Trenti, Ashby,
  Huang, Spitler, Straatman, Bradley, \& Magee}]{Bouwens2016}
Bouwens R.~J. {et~al.}, 2016, The Astrophysical Journal, 830, 67

\bibitem[{Bower {et~al}\mbox{.}(2006)Bower, Benson, Malbon, Helly, Frenk,
  Baugh, Cole, \& Lacey}]{Bower2005}
Bower R.~G., Benson A.~J., Malbon R., Helly J.~C., Frenk C.~S., Baugh C.~M.,
  Cole S., Lacey C.~G., 2006, Monthly Notices of the Royal Astronomical
  Society, 370, 645

\bibitem[{Bowler {et~al}\mbox{.}(2015)Bowler, Dunlop, McLure, McCracken,
  Milvang-Jensen, Furusawa, Taniguchi, {Le F{\`{e}}vre}, Fynbo, Jarvis, \&
  H{\"{a}}u{\ss}ler}]{Bowler2015}
Bowler R. A.~A. {et~al.}, 2015, Monthly Notices of the Royal Astronomical
  Society, 452, 1817

\bibitem[{Bowler {et~al}\mbox{.}(2017)Bowler, Dunlop, McLure, \&
  McLeod}]{Bowler2016a}
Bowler R. A.~A., Dunlop J.~S., McLure R.~J., McLeod D.~J., 2017, Monthly
  Notices of the Royal Astronomical Society, 466, 3612

\bibitem[{Bowler {et~al}\mbox{.}(2014)Bowler, Dunlop, McLure, Rogers,
  McCracken, Milvang-Jensen, Furusawa, Fynbo, Taniguchi, Afonso, Bremer, \& {Le
  Fevre}}]{Bowler2014}
Bowler R. A.~A. {et~al.}, 2014, Monthly Notices of the Royal Astronomical
  Society, 440, 2810

\bibitem[{Brainerd {et~al}\mbox{.}(1996)Brainerd, Blandford, \&
  Smail}]{Brainerd1995}
Brainerd T.~G., Blandford R.~D., Smail I., 1996, The Astrophysical Journal,
  466, 623

\bibitem[{Cabre {et~al}\mbox{.}(2007)Cabre, Fosalba, Gaztanaga, \&
  Manera}]{Cabre2007}
Cabre A., Fosalba P., Gaztanaga E., Manera M., 2007, Monthly Notices of the
  Royal Astronomical Society, 381, 1347

\bibitem[{Cen \& Safarzadeh(2014)}]{Cen2015}
Cen R., Safarzadeh M., 2014, The Astrophysical Journal, 798, L38

\bibitem[{Clay {et~al}\mbox{.}(2015)Clay, Thomas, Wilkins, \&
  Henriques}]{Clay2015}
Clay S.~J., Thomas P.~A., Wilkins S.~M., Henriques B. M.~B., 2015, Monthly
  Notices of the Royal Astronomical Society, 451, 2692

\bibitem[{Conroy {et~al}\mbox{.}(2006)Conroy, Wechsler, \&
  Kravtsov}]{Conroy2006}
Conroy C., Wechsler R.~H., Kravtsov A.~V., 2006, The Astrophysical Journal,
  647, 201

\bibitem[{Cooray \& Sheth(2002)}]{Cooray2002}
Cooray A., Sheth R., 2002, Physics Reports, 372, 1

\bibitem[{Coupon {et~al}\mbox{.}(2015)Coupon, Arnouts, van Waerbeke, Moutard,
  Ilbert, van Uitert, Erben, Garilli, Guzzo, Heymans, Hildebrandt, Hoekstra,
  Kilbinger, Kitching, Mellier, Miller, Scodeggio, Bonnett, Branchini,
  Davidzon, {De Lucia}, Fritz, Fu, Hudelot, Hudson, Kuijken, Leauthaud, {Le
  Fevre}, McCracken, Moscardini, Rowe, Schrabback, Semboloni, \&
  Velander}]{Coupon2015}
Coupon J. {et~al.}, 2015, Monthly Notices of the Royal Astronomical Society,
  449, 1352

\bibitem[{Coupon {et~al}\mbox{.}(2012)Coupon, Kilbinger, McCracken, Ilbert,
  Arnouts, Mellier, Abbas, de~la Torre, Goranova, Hudelot, Kneib, \& {Le
  F{\`{e}}vre}}]{Coupon2012}
Coupon J. {et~al.}, 2012, Astronomy {\&} Astrophysics, 542, A5

\bibitem[{Cress {et~al}\mbox{.}(1996)Cress, Helfand, Becker, Gregg, \&
  White}]{Cress1996}
Cress C.~M., Helfand D.~J., Becker R.~H., Gregg M.~D., White R.~L., 1996, The
  Astrophysical Journal, 473, 7

\bibitem[{Croton {et~al}\mbox{.}(2006)Croton, Springel, White, {De Lucia},
  Frenk, Gao, Jenkins, Kauffmann, Navarro, \& Yoshida}]{Croton2006}
Croton D.~J. {et~al.}, 2006, Monthly Notices of the Royal Astronomical Society,
  365, 11

\bibitem[{Dunlop {et~al}\mbox{.}(2013)Dunlop, Rogers, McLure, Ellis, Robertson,
  Koekemoer, Dayal, Curtis-Lake, Wild, Charlot, Bowler, Schenker, Ouchi, Ono,
  Cirasuolo, Furlanetto, Stark, Targett, \& Schneider}]{Dunlop2013}
Dunlop J.~S. {et~al.}, 2013, Monthly Notices of the Royal Astronomical Society,
  432, 3520

\bibitem[{Finkelstein {et~al}\mbox{.}(2015)Finkelstein, Ryan, Papovich,
  Dickinson, Song, Somerville, Ferguson, Salmon, Giavalisco, Koekemoer, Ashby,
  Behroozi, Castellano, Dunlop, Faber, Fazio, Fontana, Grogin, Hathi, Jaacks,
  Kocevski, Livermore, McLure, Merlin, Mobasher, Newman, Rafelski, Tilvi, \&
  Willner}]{Finkelstein2015}
Finkelstein S.~L. {et~al.}, 2015, The Astrophysical Journal, 810, 71

\bibitem[{Fisher {et~al}\mbox{.}(1994)Fisher, Davis, Strauss, Yahil, \&
  Huchra}]{Fisher1994}
Fisher J., Davis K.~B., Strauss M., Yahil M.~A., Huchra A., 1994, Monthly
  Notices of the Royal Astronomical Society, 266

\bibitem[{Foreman-Mackey {et~al}\mbox{.}(2013)Foreman-Mackey, Hogg, Lang, \&
  Goodman}]{Foreman-Mackey2012}
Foreman-Mackey D., Hogg D.~W., Lang D., Goodman J., 2013, Publications of the
  Astronomical Society of the Pacific, 125, 306

\bibitem[{Furusawa {et~al}\mbox{.}(2008)Furusawa, Kosugi, Akiyama, Takata,
  Sekiguchi, Tanaka, Iwata, Kajisawa, Yasuda, Doi, Ouchi, Simpson, Shimasaku,
  Yamada, Furusawa, Morokuma, Ishida, Aoki, Fuse, Imanishi, Iye, Karoji,
  Kobayashi, Kodama, Komiyama, Maeda, Miyazaki, Mizumoto, Nakata, Noumaru,
  Ogasawara, Okamura, Saito, Sasaki, Ueda, \& Yoshida}]{Furusawa2008}
Furusawa H. {et~al.}, 2008, The Astrophysical Journal Supplement Series, 176, 1

\bibitem[{Giavalisco(2002)}]{Giavalisco2002}
Giavalisco M., 2002, Annual Review of Astronomy and Astrophysics, 40, 579

\bibitem[{Gott {et~al}\mbox{.}(2009)Gott, Choi, Park, \& Kim}]{Gott2008a}
Gott J.~R., Choi Y.-Y., Park C., Kim J., 2009, The Astrophysical Journal, 695,
  L45

\bibitem[{Guhathakurta {et~al}\mbox{.}(1990)Guhathakurta, Tyson, \&
  Majewski}]{Guhathakurta1990}
Guhathakurta P., Tyson J.~A., Majewski S.~R., 1990, The Astrophysical Journal,
  357, L9

\bibitem[{Guo {et~al}\mbox{.}(2016)Guo, Zheng, Behroozi, Zehavi, Chuang,
  Comparat, Favole, Gottloeber, Klypin, Prada, Rodr{\'{i}}guez-Torres,
  Weinberg, \& Yepes}]{Guo2015}
Guo H. {et~al.}, 2016, Monthly Notices of the Royal Astronomical Society, 459,
  3040

\bibitem[{Harikane {et~al}\mbox{.}(2016)Harikane, Ouchi, Ono, More, Saito, Lin,
  Coupon, Shimasaku, Shibuya, Price, Lin, Hsieh, Ishigaki, Komiyama, Silverman,
  Takata, Tamazawa, \& Toshikawa}]{Harikane2015}
Harikane Y. {et~al.}, 2016, The Astrophysical Journal, 821, 123

\bibitem[{Harikane {et~al}\mbox{.}(2017)Harikane, Ouchi, Ono, Saito, Behroozi,
  More, Shimasaku, Toshikawa, Lin, Akiyama, Coupon, Komiyama, Konno, Lin,
  Miyazaki, Nishizawa, Shibuya, \& Silverman}]{Harikane2017}
Harikane Y. {et~al.}, 2017

\bibitem[{Hartley {et~al}\mbox{.}(2013)Hartley, Almaini, \&
  Foucaud}]{Hartley2013b}
Hartley W.~G., Almaini O., Foucaud S., 2013, in Thirty Years of Astronomical
  Discovery with UKIRT, Astrophysics and Space Science Proceedings, Vol.~37,
  Springer Science+Business Media, pp. 309--321

\bibitem[{Hatfield \& Jarvis(2016)}]{Hatfield2016a}
Hatfield P.~W., Jarvis M.~J., 2016, eprint arXiv:1606.08989

\bibitem[{Hatfield {et~al}\mbox{.}(2016)Hatfield, Lindsay, Jarvis,
  H{\"{a}}u{\ss}ler, Vaccari, \& Verma}]{Hatfield2016}
Hatfield P.~W., Lindsay S.~N., Jarvis M.~J., H{\"{a}}u{\ss}ler B., Vaccari M.,
  Verma A., 2016, Monthly Notices of the Royal Astronomical Society, 459, 2618

\bibitem[{Hearin {et~al}\mbox{.}(2016)Hearin, Behroozi, \& van~den
  Bosch}]{Hearin2015}
Hearin A.~P., Behroozi P.~S., van~den Bosch F.~C., 2016, Monthly Notices of the
  Royal Astronomical Society, 461, 2135

\bibitem[{H{\"{o}}nig {et~al}\mbox{.}(2017)H{\"{o}}nig, Watson, Kishimoto,
  Gandhi, Goad, Horne, Shankar, Banerji, Boulderstone, Jarvis, Smith, \&
  Sullivan}]{Honig2016}
H{\"{o}}nig S.~F. {et~al.}, 2017, Monthly Notices of the Royal Astronomical
  Society, 464, 1693

\bibitem[{Ilbert {et~al}\mbox{.}(2006)Ilbert, Arnouts, McCracken, Bolzonella,
  Bertin, {Le F{\`{e}}vre}, Mellier, Zamorani, Pell{\`{o}}, Iovino, Tresse, {Le
  Brun}, Bottini, Garilli, Maccagni, Picat, Scaramella, Scodeggio, Vettolani,
  Zanichelli, Adami, Bardelli, Cappi, Charlot, Ciliegi, Contini, Cucciati,
  Foucaud, Franzetti, Gavignaud, Guzzo, Marano, Marinoni, Mazure, Meneux,
  Merighi, Paltani, Pollo, Pozzetti, Radovich, Zucca, Bondi, Bongiorno,
  Busarello, {De La Torre}, Gregorini, Lamareille, Mathez, Merluzzi, Ripepi,
  Rizzo, \& Vergani}]{Ilbert2006}
Ilbert O. {et~al.}, 2006, Astronomy and Astrophysics, 457, 841

\bibitem[{Jarvis {et~al}\mbox{.}(2013)Jarvis, Bonfield, Bruce, Geach, McAlpine,
  McLure, Gonzalez-Solares, Irwin, Lewis, Yoldas, Andreon, Cross, Emerson,
  Dalton, Dunlop, Hodgkin, Le, Karouzos, Meisenheimer, Oliver, Rawlings,
  Simpson, Smail, Smith, Sullivan, Sutherland, White, \& Zwart}]{Jarvis2013}
Jarvis M.~J. {et~al.}, 2013, Monthly Notices of the Royal Astronomical Society,
  428, 1281

\bibitem[{Johnston(2011)}]{Johnston2011}
Johnston R., 2011, The Astronomy and Astrophysics Review, 19, 41

\bibitem[{Jose {et~al}\mbox{.}(2017)Jose, Baugh, Lacey, \&
  Subramanian}]{Jose2017}
Jose C., Baugh C.~M., Lacey C.~G., Subramanian K., 2017, eprint
  arXiv:1702.00853

\bibitem[{Jose {et~al}\mbox{.}(2016)Jose, Lacey, \& Baugh}]{Jose2016}
Jose C., Lacey C.~G., Baugh C.~M., 2016, Monthly Notices of the Royal
  Astronomical Society, 463, 270

\bibitem[{Jose {et~al}\mbox{.}(2013)Jose, Srianand, \& Subramanian}]{Jose2013}
Jose C., Srianand R., Subramanian K., 2013, Monthly Notices of the Royal
  Astronomical Society, 435, 368

\bibitem[{Jullo {et~al}\mbox{.}(2007)Jullo, Kneib, Limousin,
  El{\'{i}}asd{\'{o}}ttir, Marshall, \& Verdugo}]{Jullo2007}
Jullo E., Kneib J.-P., Limousin M., El{\'{i}}asd{\'{o}}ttir {\'{A}}., Marshall
  P.~J., Verdugo T., 2007, New Journal of Physics, 9, 447

\bibitem[{Kravtsov {et~al}\mbox{.}(2004)Kravtsov, Berlind, Wechsler, Klypin,
  Gottlober, Allgood, \& Primack}]{Kravtsov2004}
Kravtsov A.~V., Berlind A.~A., Wechsler R.~H., Klypin A.~A., Gottlober S.,
  Allgood B., Primack J.~R., 2004, The Astrophysical Journal, 609, 35

\bibitem[{Lacasa \& Rosenfeld(2016)}]{Lacasa2016}
Lacasa F., Rosenfeld R., 2016, Journal of Cosmology and Astroparticle Physics,
  2016, 005

\bibitem[{Lacey {et~al}\mbox{.}(2016)Lacey, Baugh, Frenk, Benson, Bower, Cole,
  Gonzalez-Perez, Helly, Lagos, \& Mitchell}]{Lacey2016}
Lacey C.~G. {et~al.}, 2016, Monthly Notices of the Royal Astronomical Society,
  462, 3854

\bibitem[{Laigle {et~al}\mbox{.}(2016)Laigle, McCracken, Ilbert, Hsieh,
  Davidzon, Capak, Hasinger, Silverman, Pichon, Coupon, Aussel, {Le Borgne},
  Caputi, Cassata, Chang, Civano, Dunlop, Fynbo, Kartaltepe, Koekemoer, {Le
  F{\`{e}}vre}, {Le Floc'h}, Leauthaud, Lilly, Lin, Marchesi, Milvang-Jensen,
  Salvato, Sanders, Scoville, Smolcic, Stockmann, Taniguchi, Tasca, Toft,
  Vaccari, \& Zabl}]{Laigle2016}
Laigle C. {et~al.}, 2016, The Astrophysical Journal Supplement Series, 224, 24

\bibitem[{Landy \& Szalay(1993)}]{Landy1993}
Landy S.~D., Szalay A.~S., 1993, The Astrophysical Journal, 412, 64

\bibitem[{Lawrence {et~al}\mbox{.}(2007)Lawrence, Warren, Almaini, Edge,
  Hambly, Jameson, Lucas, Casali, Adamson, Dye, Emerson, Foucaud, Hewett,
  Hirst, Hodgkin, Irwin, Lodieu, McMahon, Simpson, Smail, Mortlock, \&
  Folger}]{Lawrence2006}
Lawrence A. {et~al.}, 2007, Monthly Notices of the Royal Astronomical Society,
  379, 1599

\bibitem[{Limber(1954)}]{Limber:1954zz}
Limber D.~N., 1954, The Astrophysical Journal, 119, 655

\bibitem[{Lindsay {et~al}\mbox{.}(2014)Lindsay, Jarvis, \&
  McAlpine}]{Lindsay2014}
Lindsay S.~N., Jarvis M.~J., McAlpine K., 2014, Monthly Notices of the Royal
  Astronomical Society, 440, 2322

\bibitem[{Ling {et~al}\mbox{.}(1986)Ling, Barrow, \& Frenk}]{Ling1986}
Ling E.~N., Barrow J.~D., Frenk C.~S., 1986, Monthly Notices of the Royal
  Astronomical Society, 223, 21P

\bibitem[{Ma \& Fry(2000)}]{Ma2000}
Ma C.-P., Fry J.~N., 2000, The Astrophysical Journal, 543, 503

\bibitem[{Madau \& Dickinson(2014)}]{Madau2014}
Madau P., Dickinson M., 2014, Annual Review of Astronomy and Astrophysics, 52,
  415

\bibitem[{Maddox {et~al}\mbox{.}(1990)Maddox, Efstathiou, Sutherland, \&
  Loveday}]{Maddox1990}
Maddox S.~J., Efstathiou G., Sutherland W.~J., Loveday J., 1990, Monthly
  Notices of the Royal Astronomical Society, 242, 43P

\bibitem[{Mandelbaum {et~al}\mbox{.}(2014)Mandelbaum, Rowe, Bosch, Chang,
  Courbin, Gill, Jarvis, Kannawadi, Kacprzak, Lackner, Leauthaud, Miyatake,
  Nakajima, Rhodes, Simet, Zuntz, Armstrong, Bridle, Coupon, Dietrich, Gentile,
  Heymans, Jurling, Kent, Kirkby, Margala, Massey, Melchior, Peterson, Roodman,
  \& Schrabback}]{Mandelbaum2013}
Mandelbaum R. {et~al.}, 2014, The Astrophysical Journal Supplement Series, 212,
  5

\bibitem[{Mann {et~al}\mbox{.}(1998)Mann, Peacock, \& Heavens}]{Mann1997a}
Mann R.~G., Peacock J.~A., Heavens A.~F., 1998, Monthly Notices of the Royal
  Astronomical Society, 293, 209

\bibitem[{McCracken {et~al}\mbox{.}(2012)McCracken, Milvang-Jensen, Dunlop,
  Franx, Fynbo, {Le F{\`{e}}vre}, Holt, Caputi, Goranova, Buitrago, Emerson,
  Freudling, Hudelot, L{\'{o}}pez-Sanjuan, Magnard, Mellier, M{\o}ller,
  Nilsson, Sutherland, Tasca, \& Zabl}]{McCracken2012}
McCracken H.~J. {et~al.}, 2012, Astronomy {\&} Astrophysics, 544, A156

\bibitem[{McCracken {et~al}\mbox{.}(2015)McCracken, Wolk, Colombi, Kilbinger,
  Ilbert, Peirani, Coupon, Dunlop, Milvang-Jensen, Caputi, Aussel, Bethermin,
  \& {Le Fevre}}]{McCracken2015}
McCracken H.~J. {et~al.}, 2015, Monthly Notices of the Royal Astronomical
  Society, 449, 901

\bibitem[{McLeod {et~al}\mbox{.}(2016)McLeod, McLure, \& Dunlop}]{McLeod2016}
McLeod D.~J., McLure R.~J., Dunlop J.~S., 2016, Monthly Notices of the Royal
  Astronomical Society, 459, 3812

\bibitem[{McLure {et~al}\mbox{.}(2009)McLure, Cirasuolo, Dunlop, Foucaud, \&
  Almaini}]{McLure2009}
McLure R.~J., Cirasuolo M., Dunlop J.~S., Foucaud S., Almaini O., 2009, Monthly
  Notices of the Royal Astronomical Society, 395, 2196

\bibitem[{McLure {et~al}\mbox{.}(2013)McLure, Dunlop, Bowler, Curtis-Lake,
  Schenker, Ellis, Robertson, Koekemoer, Rogers, Ono, Ouchi, Charlot, Wild,
  Stark, Furlanetto, Cirasuolo, \& Targett}]{McLure2013}
McLure R.~J. {et~al.}, 2013, Monthly Notices of the Royal Astronomical Society,
  432, 2696

\bibitem[{McQuinn {et~al}\mbox{.}(2007)McQuinn, Hernquist, Zaldarriaga, \&
  Dutta}]{McQuinn2007}
McQuinn M., Hernquist L., Zaldarriaga M., Dutta S., 2007, Monthly Notices of
  the Royal Astronomical Society, 381, 75

\bibitem[{Mo {et~al}\mbox{.}(1992)Mo, Jing, \& Boerner}]{Mo1992}
Mo H.~J., Jing Y.~P., Boerner G., 1992, The Astrophysical Journal, 392, 452

\bibitem[{Moster {et~al}\mbox{.}(2010)Moster, Somerville, Maulbetsch, van~den
  Bosch, Macci{\`{o}}, Naab, \& Oser}]{Moster2009}
Moster B.~P., Somerville R.~S., Maulbetsch C., van~den Bosch F.~C.,
  Macci{\`{o}} A.~V., Naab T., Oser L., 2010, The Astrophysical Journal, 710,
  903

\bibitem[{Murray {et~al}\mbox{.}(2013)Murray, Power, \& Robotham}]{Murray2013}
Murray S.~G., Power C., Robotham A. S.~G., 2013, Monthly Notices of the Royal
  Astronomical Society: Letters, 434, L61

\bibitem[{Natarajan \& Yoshida(2014)}]{Natarajan2014}
Natarajan A., Yoshida N., 2014, Progress of Theoretical and Experimental
  Physics, 6B112

\bibitem[{Navarro {et~al}\mbox{.}(1996)Navarro, Frenk, \& White}]{Navarro1996}
Navarro J.~F., Frenk C.~S., White S. D.~M., 1996, The Astrophysical Journal,
  462, 563

\bibitem[{Newman \& Davis(2002)}]{Newman2001}
Newman J.~A., Davis M., 2002, The Astrophysical Journal, 564, 567

\bibitem[{Norberg {et~al}\mbox{.}(2009)Norberg, Baugh, Gazta{\~{n}}aga, \&
  Croton}]{Norberg2008}
Norberg P., Baugh C.~M., Gazta{\~{n}}aga E., Croton D.~J., 2009, Monthly
  Notices of the Royal Astronomical Society, 396, 19

\bibitem[{Oesch {et~al}\mbox{.}(2016)Oesch, Brammer, van Dokkum, Illingworth,
  Bouwens, Labb{\'{e}}, Franx, Momcheva, Ashby, Fazio, Gonzalez, Holden, Magee,
  Skelton, Smit, Spitler, Trenti, \& Willner}]{Oesch2016}
Oesch P.~A. {et~al.}, 2016, The Astrophysical Journal, 819, 129

\bibitem[{Oke \& Gunn(1983)}]{Oke1983}
Oke J.~B., Gunn J.~E., 1983, The Astrophysical Journal, 266, 713

\bibitem[{Ono {et~al}\mbox{.}(2017)Ono, Ouchi, Harikane, Toshikawa, Rauch,
  Yuma, Sawicki, Shibuya, Shimasaku, Oguri, Willott, Akhlaghi, Akiyama, Coupon,
  Kashikawa, Komiyama, Konno, Lin, Matsuoka, Miyazaki, Nagao, Nakajima,
  Silverman, Tanaka, \& Wang}]{Ono2017}
Ono Y. {et~al.}, 2017

\bibitem[{Ouchi {et~al}\mbox{.}(2009)Ouchi, Mobasher, Shimasaku, Ferguson,
  Fall, Ono, Kashikawa, Morokuma, Nakajima, Okamura, Dickinson, Giavalisco, \&
  Ohta}]{Ouchi2009}
Ouchi M. {et~al.}, 2009, The Astrophysical Journal, 706, 1136

\bibitem[{Ouchi {et~al}\mbox{.}(2010)Ouchi, Shimasaku, Furusawa, Saito,
  Yoshida, Akiyama, Ono, Yamada, Ota, Kashikawa, Iye, Kodama, Okamura, Simpson,
  \& Yoshida}]{Ouchi2010}
Ouchi M. {et~al.}, 2010, The Astrophysical Journal, 723, 869

\bibitem[{Paczynski(1987)}]{Paczynski1987}
Paczynski B., 1987, Nature, 325, 572

\bibitem[{Peebles(1980)}]{Peebles1980}
Peebles P. J.~E., 1980, {The large-scale structure of the universe}. Princeton
  University Press

\bibitem[{Peebles(1993)}]{Peebles1993}
Peebles P. J.~E., 1993, {Principles of physical cosmology}. Princeton
  University Press

\bibitem[{Peebles \& Hauser(1974)}]{Peebles1974}
Peebles P. J.~E., Hauser M.~G., 1974, The Astrophysical Journal Supplement
  Series, 28, 19

\bibitem[{Pentericci {et~al}\mbox{.}(2014)Pentericci, Vanzella, Fontana,
  Castellano, Treu, Mesinger, Dijkstra, Grazian, Brada{\v{c}}, Conselice,
  Cristiani, Dunlop, Galametz, Giavalisco, Giallongo, Koekemoer, McLure,
  Maiolino, Paris, \& Santini}]{Pentericci2014}
Pentericci L. {et~al.}, 2014, The Astrophysical Journal, 793, 113

\bibitem[{Roche \& Eales(1999)}]{Roche1999}
Roche N., Eales S.~A., 1999, Monthly Notices of the Royal Astronomical Society,
  307, 703

\bibitem[{Schechter(1976)}]{Schechter1976}
Schechter P., 1976, The Astrophysical Journal, 203, 297

\bibitem[{Scoville {et~al}\mbox{.}(2007)Scoville, Abraham, Aussel, Barnes,
  Benson, Blain, Calzetti, Comastri, Capak, Carilli, Carlstrom, Carollo,
  Colbert, Daddi, Ellis, Elvis, Ewald, Fall, Franceschini, Giavalisco, Green,
  Griffiths, Guzzo, Hasinger, Impey, Kneib, Koda, Koekemoer, Lefevre, Lilly,
  Liu, McCracken, Massey, Mellier, Miyazaki, Mobasher, Mould, Norman,
  Refregier, Renzini, Rhodes, Rich, Sanders, Schiminovich, Schinnerer,
  Scodeggio, Sheth, Shopbell, Taniguchi, Tyson, Urry, {Van Waerbeke},
  Vettolani, White, \& Yan}]{Scoville2007}
Scoville N. {et~al.}, 2007, The Astrophysical Journal Supplement Series, 172,
  38

\bibitem[{Seljak(2000)}]{Seljak2000}
Seljak U., 2000, Monthly Notices of the Royal Astronomical Society, 318, 203

\bibitem[{Shao(1986)}]{Shao1986}
Shao J., 1986, The Annals of Statistics, 14, 1322

\bibitem[{Shapley(2011)}]{Shapley2011}
Shapley A.~E., 2011, Annual Review of Astronomy and Astrophysics, 49, 525

\bibitem[{Silk(2010)}]{Silk2011}
Silk J., 2010, Proceedings of the International Astronomical Union, 6, 273

\bibitem[{Somerville {et~al}\mbox{.}(2004)Somerville, Lee, Ferguson, Gardner,
  Moustakas, \& Giavalisco}]{Somerville2003}
Somerville R.~S., Lee K., Ferguson H.~C., Gardner J.~P., Moustakas L.~A.,
  Giavalisco M., 2004, The Astrophysical Journal, 600, L171

\bibitem[{Soucail {et~al}\mbox{.}(1988)Soucail, Mellier, Fort, Mathez, \&
  Cailloux}]{Soucail1988}
Soucail G., Mellier Y., Fort B., Mathez G., Cailloux M., 1988, {The giant arc
  in A 370 - Spectroscopic evidence for gravitational lensing from a source at
  Z = 0.724}, Vol. 191. EDP Sciences [etc.], pp. L19--L21

\bibitem[{Stark(2016)}]{Stark2016}
Stark D.~P., 2016, Annual Review of Astronomy and Astrophysics, 54, 761

\bibitem[{Stark {et~al}\mbox{.}(2009)Stark, Ellis, Bunker, Bundy, Targett,
  Benson, \& Lacy}]{Stark2009}
Stark D.~P., Ellis R.~S., Bunker A., Bundy K., Targett T., Benson A., Lacy M.,
  2009, The Astrophysical Journal, Volume 697, Issue 2, pp. 1493-1511 (2009).,
  697, 1493

\bibitem[{Steidel {et~al}\mbox{.}(1996)Steidel, Giavalisco, Pettini, Dickinson,
  \& Adelberger}]{Steidel1996}
Steidel C.~C., Giavalisco M., Pettini M., Dickinson M., Adelberger K.~L., 1996,
  The Astrophysical Journal, 462, L17

\bibitem[{Steidel \& Hamilton(1992)}]{Steidel1992}
Steidel C.~C., Hamilton D., 1992, The Astronomical Journal, 104, 941

\bibitem[{Tegmark \& Max(2003)}]{Tegmark2003}
Tegmark M., Max, 2003, American Physical Society, April Meeting, 2003, April
  5-8, 2003 Philadelphia, Pennsylvania, MEETING ID: APR03, abstract {\#}T1.003

\bibitem[{Tinker {et~al}\mbox{.}(2008)Tinker, Kravtsov, Klypin, Abazajian,
  Warren, Yepes, Gottl{\"{o}}ber, \& Holz}]{Tinker2008}
Tinker J., Kravtsov A.~V., Klypin A., Abazajian K., Warren M., Yepes G.,
  Gottl{\"{o}}ber S., Holz D.~E., 2008, The Astrophysical Journal, 688, 709

\bibitem[{Tinker {et~al}\mbox{.}(2010)Tinker, Robertson, Kravtsov, Klypin,
  Warren, Yepes, \& Gottl{\"{o}}ber}]{Tinker2010}
Tinker J.~L., Robertson B.~E., Kravtsov A.~V., Klypin A., Warren M.~S., Yepes
  G., Gottl{\"{o}}ber S., 2010, The Astrophysical Journal, 724, 878

\bibitem[{Trenti \& Stiavelli(2008)}]{Trenti2008}
Trenti M., Stiavelli M., 2008, The Astrophysical Journal, 676, 767

\bibitem[{Vale \& Ostriker(2004)}]{Vale2004}
Vale A., Ostriker J.~P., 2004, Monthly Notices of the Royal Astronomical
  Society, 353, 189

\bibitem[{White(1979)}]{White1979}
White S. D.~M., 1979, Monthly Notices of the Royal Astronomical Society, 186,
  145

\bibitem[{Zaroubi(2013)}]{Zaroubi2013}
Zaroubi S., 2013, in The First Galaxies, Astrophysics and Space Science
  Library, Vol. 396, Springer, pp. 45--101

\bibitem[{Zehavi {et~al}\mbox{.}(2005)Zehavi, Eisenstein, Nichol, Blanton,
  Hogg, Brinkmann, Loveday, Meiksin, Schneider, \& Tegmark}]{Zehavi2005}
Zehavi I. {et~al.}, 2005, The Astrophysical Journal, 621, 22

\bibitem[{Zheng {et~al}\mbox{.}(2005)Zheng, Berlind, Weinberg, Benson, Baugh,
  Cole, Dave, Frenk, Katz, \& Lacey}]{Zheng2005}
Zheng Z. {et~al.}, 2005, The Astrophysical Journal, 633, 791

\end{thebibliography}

\bsp

\label{lastpage}

\end{document}